\documentclass[journal]{IEEEtran}

\usepackage[T1]{fontenc}
\usepackage{textcomp}
\usepackage{blindtext}
\usepackage[utf8]{inputenc}
\usepackage[noadjust]{cite}
\usepackage[font=footnotesize]{caption}
\usepackage{dblfloatfix}
\usepackage{graphicx}
\usepackage{psfrag}
\usepackage{subfig}
\usepackage{amsmath,amssymb,amsthm,mathrsfs,amsfonts,dsfont}
\usepackage{color}
\usepackage[algoruled]{algorithm2e}
\usepackage{fixltx2e}
\usepackage{multirow}
\usepackage{multicol}
\usepackage[hidelinks]{hyperref}
\usepackage[normalem]{ulem}
\usepackage{acronym}
\usepackage[table,xcdraw]{xcolor}
\usepackage{accents}
\usepackage{mwe}
\usepackage{hhline}
\usepackage{trfsigns}
\usepackage{siunitx}
\usepackage{lscape}
\usepackage{wasysym}

\DeclareMathOperator*{\argmin}{arg\,min}

\theoremstyle{definition}

\newtheorem{thm_rem}{Remark}

\newacro{5G}{fifth generation}
\newacro{6G}{sixth generation}
\newacro{A/D}{analog-to-digital}
\newacro{ADC}{analog-to-digital converter}
\newacro{AFE}{analog front-end}
\newacro{AGV}{automatic guided vehicle}
\newacro{AWGN}{additive white Gaussian noise}
\newacro{B5G}{beyond \ac{5G}}
\newacro{BB}{baseband}
\newacro{BER}{bit error ratio}
\newacro{BPSK}{binary phase-shift keying}
\newacro{BP}{band-pass}
\newacro{BS}{base station}
\newacro{CDM}{code-division multiplexing}
\newacro{CFO}{carrier frequency offset}
\newacro{CFR}{channel frequency response}
\newacro{CIR}{channel impulse response}
\newacro{CoMP}{coordinated multipoint}
\newacro{CP}{cyclic prefix}
\newacro{CPO}{carrier phase offset}
\newacro{CRLB}{Cram\'er–Rao lower bound}
\newacro{CS}{chirp sequence}
\newacro{CW}{continuous wave}
\newacro{CZT}{chirp Z-transform}
\newacro{D/A}{digital-to-analog}
\newacro{DAC}{digital-to-analog converter}
\newacro{DDS}{direct digital synthesis}
\newacro{DFRC}{dual-function radar-cunication or dual-functional radar-cunication}
\newacro{DFnT}{discrete Fresnel transform}
\newacro{DFT}{discrete Fourier transform}
\newacro{DMRS}{demodulation reference signal}
\newacro{EVM}{error vector magnitude}
\newacro{FDE}{frequency-domain equalization}
\newacro{FDM}{frequency-division multiplexing}
\newacro{FO}{frequency offset}
\newacro{gNB}{gNodeB}
\newacro{HP}{high-pass}
\newacro{ICI}{intercarrier interference}
\newacro{IDFT}{inverse discrete Fourier transform}
\newacro{IDFnT}{inverse discrete Fresnel transform}
\newacro{IHE}{Institute of Radio Frequency Engineering and Electronics}
\newacro{ISAC}{integrated sensing and communication}
\newacro{ISI}{intersymbol interference}
\newacro{ISLR}{integrated-sidelobe level ratio}
\newacro{JCAS}{joint communication and sensing}
\newacro{KIT}{Karlsruhe Institute of Technology}
\newacro{LDPC}{low-density parity-check}
\newacro{LFSR}{linear-feedback shift register}
\newacro{LNA}{low-noise amplifier}
\newacro{LoS}{line-of-sight}
\newacro{LP}{low-pass}
\newacro{LS}{least squares}
\newacro{mmWave}{milimeter wave}
\newacro{MIMO}{multiple-input multiple-output}
\newacro{MLE}{maximum likelihood estimator}
\newacro{MLS}{maximum-length sequence}
\newacro{MUSIC}{multiple signal classification}
\newacro{NLoS}{non-line-of-sight}
\newacro{NR}{new radio}
\newacro{OCDM}{orthogonal chirp-division multiplexing}
\newacro{OFDM}{orthogonal frequency-division multiplexing}
\newacro{OOB}{out-of-band}
\newacro{OTA}{over-the-air}
\newacro{P/S}{parallel-to-serial}
\newacro{PA}{power amplifier}
\newacro{PACF}{periodic autocorrelation function}
\newacro{PCCF}{periodic cross-correlation function}
\newacro{PLC}{powerline cunication}
\newacro{PLL}{phase-locked loop}
\newacro{PMCW}{phase-modulated continuous wave}
\newacro{PMN}{perceptive mobile network}
\newacro{PoC}{proof-of-concept}
\newacro{PRBS}{pseudorandom binary sequence}
\newacro{PRS}{positioning reference signal}
\newacro{PSLR}{peak-to-sidelobe level ratio}
\newacro{QPSK}{quadrature phase-shift keying}
\newacro{RadCom}{radar-cunication}
\newacro{RCS}{radar cross section}
\newacro{RF}{radio-frequency}
\newacro{RIS}{reflective intelligent surface}
\newacro{RMSE}{root mean squared error}
\newacro{SC}[S\&C]{Schmidl \& Cox}
\newacro{SFO}{sampling frequency offset}
\newacro{SIC}{self-interference cancellation}
\newacro{SINR}{signal-to-interference-plus-noise ratio}
\newacro{SISO}{single-input single-output}
\newacro{SNR}{signal-to-noise ratio}
\newacro{SoC}{system-on-a-chip}
\newacro{STO}{symbol time offset}
\newacro{S/P}{serial-to-parallel}
\newacro{TDE}{time-domain equalization}
\newacro{TDM}{time-division multiplexing}
\newacro{TDR}{time-domain reflectometry}
\newacro{TITO}{tilt inference of time offset}
\newacro{TO}{time offset}
\newacro{UE}{user equipment}
\newacro{UWAC}{underwater acoustic cunication}
\newacro{V2V}{vehicle-to-vehicle}
\newacro{ZF}{zero forcing}
\newacro{ZP}{zero padding}
\usepackage{pgfplots}
\pgfplotsset{compat=newest}
\usetikzlibrary{plotmarks}
\usetikzlibrary{arrows.meta}
\usepgfplotslibrary{patchplots}
\usepackage{grffile}
\usepackage{amsmath}

\usepackage{calrsfs}

\usepackage{fancyhdr}
\fancypagestyle{empty}{fancy}

\begin{document}
	
	\title{Pilot-Based SFO Estimation for Bistatic\\ Integrated Sensing and Communication}
	
	\author{Lucas Giroto de Oliveira,~\IEEEmembership{Graduate Student Member,~IEEE},
		Yueheng Li,	Silvio Mandelli,~\IEEEmembership{Member,~IEEE},\\ David Brunner, Marcus Henninger,~\IEEEmembership{Member,~IEEE}, Xiang Wan,~\IEEEmembership{Senior Member,~IEEE},\\ Tie Jun Cui,~\IEEEmembership{Fellow,~IEEE}, Thomas Zwick,~\IEEEmembership{Fellow,~IEEE}, and Benjamin Nuss,~\IEEEmembership{Senior Member,~IEEE}
		\thanks{Manuscript received DD MM, 2024. The authors acknowledge the financial support by the Federal Ministry of Education and Research of Germany in the projects ``KOMSENS-6G'' (grant number: 16KISK123) and ``Open6GHub'' (grant number: 16KISK010). \textit{(Corresponding author: Lucas Giroto de Oliveira.)}}
		\thanks{L. Giroto de Oliveira, Y. Li, T. Zwick, and B. Nuss are with the Institute of Radio Frequency Engineering and Electronics (IHE), Karlsruhe Institute of Technology (KIT), 76131 Karlsruhe, Germany (e-mail: {lucas.oliveira@kit.edu}, {yueheng.li@kit.edu}, {thomas.zwick@kit.edu}, {benjamin.nuss@kit.edu}).}
		\thanks{S. Mandelli and M. Henninger are with Nokia Bell Laboratories, 70469 Stuttgart, Germany (e-mail: {silvio.mandelli@nokia-bell-labs.com}, {marcus.henninger@nokia.com}).}
		\thanks{D. Brunner was with the Institute of Radio Frequency Engineering and Electronics (IHE), Karlsruhe Institute of Technology (KIT), 76131 Karlsruhe, Germany. He is now with VEGA Grieshaber KG, 77761 Schiltach, Germany (e-mail: {d.brunner@vega.com}).}
		\thanks{X. Wan and T. J. Cui  are with Southeast University, Nanjing, Jiangsu 12579, China (e-mail: {wan\_xiang@seu.edu.cn}, {tjcui@seu.edu.cn}).}
	}
	
	
	\maketitle
	
	\begin{abstract}
		Enabling bistatic radar sensing within the context of integrated sensing and communication (ISAC) for future sixth generation mobile networks demands strict synchronization accuracy, which is particularly challenging to be achieved with over-the-air synchronization. Existing algorithms handle time and frequency offsets adequately, but provide insufficiently accurate sampling frequency offset (SFO) estimates that result in degradation of obtained radar images in the form of signal-to-noise ratio loss and migration of range and Doppler shift. This article introduces an SFO estimation algorithm named tilt inference of time offset (TITO) for orthogonal frequency-division multiplexing (OFDM)-based ISAC. Using available pilot subcarriers, TITO obtains channel impulse response estimates and extracts information on the SFO-induced delay migration to a dominant reference path with constant range, Doppler shift, and angle between transmit and receive ISAC nodes. TITO then adaptively selects the delay estimates that are only negligibly impaired by SFO-induced intersymbol interference, ultimately employing them to estimate the SFO. Assuming a scenario without a direct line-of-sight (LoS) between the aforementioned transmitting and receiving ISAC nodes, a system concept with a relay reflective intelligent surface (RIS) is used to create the aforementioned reference path is proposed. Besides a mathematical derivation of accuracy bounds, simulation and measurements at $\SI{26.2}{\giga\hertz}$ are presented to demonstrate TITO's superiority over existing methods in terms of SFO estimation accuracy and robustness.
	\end{abstract}
	
	
	\IEEEpeerreviewmaketitle
	

	\section{Introduction}\label{sec:introduction}

	\IEEEPARstart{S}{ixth} generation (6G) networks are expected to be introduced with the aim of not only increasing connection rate and its reliability, but also enabling a multitude of functionalities beyond data communication \cite{viswanathan2020}. Among the latter, \ac{ISAC} \cite{lima2021,wild2021,liu2022,mandelli2023survey} comes as a disruptive feature with the potential of turning cellular networks into perceptive mobile networks~\cite{zhang2021}. This is expected to enable applications such as monitoring of areas with campus networks in intralogistics scenarios with humans and \acp{AGV} as well as environment sensing for critical infrastructure or safe urban mobility \cite{kadelka2023,shatov2024}.
	
	In monostatic \ac{ISAC} setups where transmit and receiver are co-located, requirements related to full-duplex operation such as high isolation between transmit and receive antenna arrays and \ac{SIC} are imposed \cite{barneto2019,barneto2021}. If, however, the transmitter and receiver are not co-located but rather widely separated, a bistatic \ac{ISAC} setup is established \cite{ksiezyk2023}. In this case, the aforementioned full-duplex requirements are avoided, and existing deployments of cellular networks are exploited, achieving greater diversity of radar sensing measurements \cite{thomae2023,bauhofer2023}. In spite of its advantages, bistatic sensing comes at the cost of strict synchronization requirements to avoid biases in radar target parameter estimates such as bistatic range and Doppler shifts. When this is not achievable on hardware-level, e.g., via optical fiber connection between sensing gNodeBs (gNBs), \ac{OTA} synchronization must be performed.

	
	In the context of bistatic sensing with cellular networks, recent studies in the literture have investigated bistatic radar sensing with \ac{5G} networks. Examples include \cite{samczynski2022,abratkiewicz2023,ksiezyk2023}, which consider bistatic sensing with a gNB acting as an illuminator of opportunity and a receiver having two separate antennas and \ac{RF}/\ac{BB} channels for the reference and sensing links. In a previous study \cite{giroto2023_EuMW}, a \ac{OFDM}-based bistatic \ac{ISAC} system concept was proposed. Unlike the passive case, this concept is based on actively beamforming the a single transmit signal from a transmitting gNB towards both the receiving gNB and the directions of radar targets. At the receiver side, beams are pointed in the direction of the transmitting gNB, creating a reference path for sensing, as well as in the illuminated directions by the transmitting gNB where potential targets are located. The aforementioned reference \ac{LoS} between is used for synchronization and communication, allowing to estimate the full transmit \ac{OFDM} frame and use it for radar signal processing \cite{giroto2021_tmtt} to generate a bistatic radar image.
	
	It was observed in \cite{giroto2023_EuMW} that small inaccuracies can be tolerated for time and frequency synchronization without biasing the radar target parameter estimates, as it is known that the \ac{LoS} path should appear at null relative bistatic range and Doppler shift in the radar image and slight offsets can be therefore recognized. The \ac{SFO}, however, must be accurately estimated to avoid range and Doppler shift migration. These effects, which were also observed in and \cite{werbunat2024}, result in \ac{SNR} and resolution loss. It was also shown in \cite{giroto2023_EuMW} that the accuracy yielded by the algorithm by Tsai et al.~\cite{tsai2005} was not enough for unbiased bistatic sensing, and the discrete-time domain shifting of the \ac{OFDM} symbols as described in \cite{burmeister2021} was adopted to correct the residual \ac{SFO}. While such a measure is important as the \ac{SFO} effects accumulate with increasing \ac{OFDM} symbol index, which is particularly critical in \ac{ISAC} applications where long frames are needed to achieve fine Doppler shift resolution and high processing gain, the aforementioned residual correction only compensates range migration, being ineffective against the \ac{SFO}-induced Doppler shift migration.
		
	An alternative delay-based \ac{SFO} estimation approach was proposed by Wu et al. in \cite{wu2012}. In this approach, a correlation between the amplitudes of consecutive \ac{CIR} estimates obtained based on pilot subcarriers is performed to estimate the \ac{SFO}-induced delay migration and, as a next step, the \ac{SFO} itself. While it is claimed in \cite{dantas2016} that this approach might achieve inaccurate estimates depending on the adopted \ac{OFDM} signal parameters, the proposed method by Wu et al. benefits from a \ac{DFT} processing gain proportional to the number of pilot subcarriers used to estimate each \ac{CIR} \cite{giroto2023_EuMW}. This enhances the effective \ac{SNR} for \ac{SFO} estimation when compared to the phase-based method by Tsai et al.~\cite{tsai2005} that operates on the subcarrier level in the discrete-frequency domain. In addition, the \ac{CIR} correlation method was proven to be effective in practice in \cite{pegoraro2024}. The latter study, however, did not explicitly mention \ac{SFO} in spite of mentioning the use of a Farrow filter \cite{farrow1988} to correct non-ideal sampling, and it only dealt with the resulting range migration caused by residual \ac{SFO}, not compensating the \ac{SFO}-induced Doppler shift migration.
	
	In this context, this article investigates an accurate pilot-based \ac{SFO} estimation to enable further correction. This enables avoiding \ac{SFO}-induced impairments to bistatic sensing in \ac{OFDM}-based \ac{ISAC} systems, namely \ac{SNR} degradation and migration of range and Doppler shift. Assuming that the reference path is much stronger than the reflections off radar targets, delay-based estimation based on the method by Wu et. al is performed and analyzed. This method consists of estimating the delay migration of the reference path along consecutive \ac{CIR} estimates obtained with \ac{OFDM} symbols containing pilots, and then calculating its slope and converting it into an \ac{SFO} estimate. While it does not profit from the contribution of target reflections to the effective \ac{SNR} for \ac{SFO} estimation, which in the assumed scenario is rather limited, it allows skipping the \ac{CIR} correlation processing and therefore reducing computational complexity. This method, however, is susceptible to impairments caused by \ac{SFO}-induced \ac{ICI} and \ac{ISI}, especially the latter one. Consequently, bias in the obtained \ac{SFO} estimates cannot be ruled out.
	
	To simultaneously ensure higher effective \ac{SNR} for \ac{SFO} estimation than in the method by Tsai et al. and avoid bias due to \ac{SFO}-induced \ac{ISI} in the method by Wu et al., therefore enabling robust communication and bistatic sensing, this article introduces the \ac{TITO} method. \ac{TITO} performs the same estimation of the delay migration slope across consecutive \ac{CIR} estimates to obtain an \ac{SFO} estimation as in \cite{wu2012}, but with an adaptive choice of the number of pilot \ac{OFDM} symbols used to obtain \ac{CIR} estimates. This complies with the flexible and irregular structure of 5G frame structures, where pilots might be available or not depending on the resource management decisions taken. In addition, it allows avoiding \ac{ISI}-contamined delay migration estimation, which would ultimately bias the obtained \ac{SFO} estimate.
	
	In addition to the proposed \ac{TITO} method for \ac{SFO} estimation, this article also addresses the challenge of achieving a dominant path with constant range, Doppler shift, and angle in an \ac{NLoS} scenario. This is particularly important, e.g., in indoor deployments or when at least one of the gNBs constituting the bistatic pair is at street level, as no \ac{LoS} is available in such cases. To this end, a system concept is proposed in which a reference path is assumed to be created via a relay \ac{RIS} \cite{bjornsson2022,chepuri2023,elbir2023}, which redirects the incoming signal from the transmitting to the receiving \ac{ISAC} node.
	
	\begin{figure*}[!t]
		\centering
		
		\psfrag{A}[c][c]{\footnotesize $R^\text{Tx---T}_p$}
		\psfrag{B}[c][c]{\footnotesize $R^\text{T---Rx}_p$}
		\psfrag{C}[c][c]{\footnotesize $R_0$}
		\psfrag{D}[c][c]{\footnotesize $x[s]$}
		\psfrag{E}[c][c]{\footnotesize $y[s]$}
		\psfrag{F}[c][c]{\footnotesize $F_\mathrm{s}^\mathrm{Tx}$}
		\psfrag{G}[c][c]{\footnotesize $F_\mathrm{s}^\mathrm{Rx}$}
		\includegraphics[height=6cm]{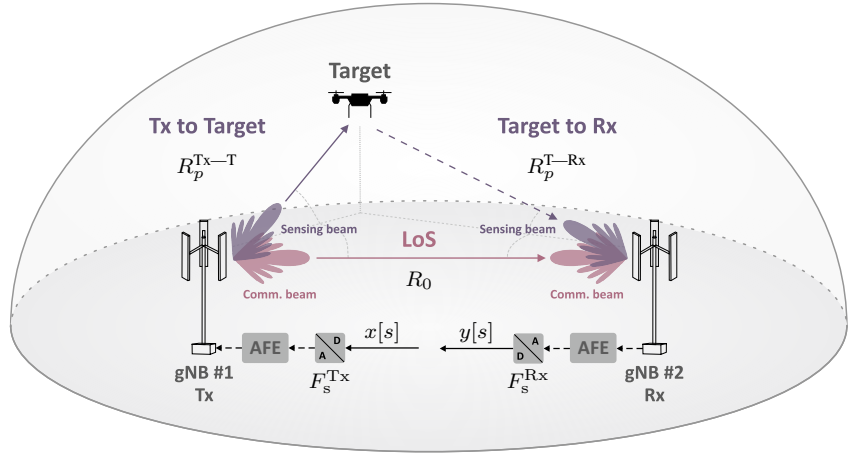}
		\captionsetup{justification=raggedright,labelsep=period,singlelinecheck=false}
		\caption{\ Bistatic ISAC system representation.}\label{fig:systemModel_gNB}	
		
	\end{figure*}

	\begin{figure*}[!b]
		\hrulefill
		\vspace*{4pt}
		\begin{equation}\label{eq:yt}
			y(t) =~\alpha_0~x(t-\tau_0-\tau_\Delta)~\e^{\im 2\pi f_\Delta t + \psi_\Delta}+\sum_{p=1}^{P-1}\alpha_p~ x(t-\tau_p-\tau_\Delta)~\e^{\im 2\pi f_{\mathrm{D},p}}~\e^{\im 2\pi f_\Delta t + \psi_\Delta}
		\end{equation}
	\end{figure*}
		
	The contributions of this article can be summarized as follows:
	\begin{itemize}
		
		\item A detailed analysis of the effects of \ac{SFO}, showing that it results in amplitude modulation of \ac{OFDM} subcarriers, besides delay and frequency shift migration that may lead to relevant \ac{ISI} and \ac{ICI}, respectively. Finally, an approximate \ac{CRLB} of the \ac{SFO} estimation is analytically derived.
		
		\item To enhance estimation accuracy in spite of \ac{SFO}-induced impairments, the \ac{TITO} method is proposed, which is by-design robust against \ac{ISI}. \ac{TITO} also potentially requires lower computational complexity than \cite{wu2012} in cases with high \ac{SFO}, as an adaptive number of pilot \ac{OFDM} symbols is used.
		
		\item A comprehensive numerical analysis to support the claims and contributions of this article, illustrating the effect of the aforementioned \ac{SFO}-induced impairments on synchronization, communication and sensing performances.
		
		\item A novel system concept for bistatic \ac{ISAC} scenarios is proposed and demonstrated in practice. It consists of a transmitter and a receiver, both performing \ac{RIS}-assisted beamsteering, besides a relay \ac{RIS} to establish a static reference path when transmitter and receiver are not in \ac{LoS}. This path is used to enable synchronization and communication, besides serving as reference to estimate range and Doppler of radar targets.
		
		\item Measurement results with the \ac{RIS}-based setup to demonstrate the superiority of the \ac{TITO} method over alternative \ac{SFO} estimation approaches from the literature.
		
	\end{itemize}
	
	The remainder of this article is organized as follows. Section~\ref{sec:system_model} presents the proposed \ac{RIS}-assisted bistatic \ac{ISAC} system concept assuming \ac{OFDM} as a modulation scheme. Next, Section~\ref{sec:SFO_sec} describes the effects of \ac{SFO} on the receive \ac{OFDM} signal and formulates both the delay migration-based \ac{SFO} estimation approach derived from the method by Wu et al.~\cite{wu2012} and the proposed \ac{TITO} method, also deriving accuracy bounds and discussing \ac{SFO} correction approaches. Section~\ref{sec:simResults} then presents a numerical analysis of the \ac{SFO} effects on both communication and sensing performances, besides analyzing the \ac{SFO} estimation performance of the \ac{TITO} method. Finally, measurement results are presented in Section~\ref{sec:measResults}, and concluding remarks are given in Section~\ref{sec:conclusion}.

	\section{System Model}\label{sec:system_model}
		
	In the considered bistatic \ac{ISAC} system, which is depicted in Fig.~\ref{fig:systemModel_gNB}, two \acp{gNB} are present. The first, \ac{gNB} \#1, acts as a transmitter, while the second, \ac{gNB} \#2, operates in the receiving mode. It is also assumed that there are $P$ propagation paths between the \acp{gNB} labeled as $p\in\{0,1,\dots,P-1\}$. The first, $p=0$, is a reference path created by the \acp{gNB} via beamforming, which enables synchronization and communication, besides providing a benchmark for bistatic radar sensing following the framework from \cite{giroto2023_EuMW}. This path has length $R_0$ that results in a propagation delay \mbox{$\tau_0 = R_0/c_0$}, where $c_0$ is the speed of light in vacuum, and Doppler shift \mbox{$f_{\mathrm{D},0}=\SI{0}{\hertz}$} since the \acp{gNB} are static. In Fig.~\ref{fig:systemModel_gNB}, this reference is assumed to be a \ac{LoS} path. Beams are also pointed from both \ac{gNB} \#1 and \ac{gNB} \#2 towards point targets that are associated with the remaining $P-1$ propagation paths. It is assumed that the $p\mathrm{th}$ path, $p\in\{1,\dots,P-1\}$, has length \mbox{$R_p=R^\text{Tx---T}_p+R^\text{T---Rx}_p$}, where $R^\text{Tx---T}_p$ and $R^\text{T---Rx}_p$ are the ranges of the associated point target with this path w.r.t. \ac{gNB} \#1 and \ac{gNB} \#2, respectively. Due to the propagation through this path, a delay \mbox{$\tau_p = R_p/c_0$} is experienced. In addition, the movement of the point target associated with the $p\mathrm{th}$ path results in a Doppler shift \mbox{$f_{\mathrm{D},p} = (2v_p/\lambda_0)~\cos(\gamma_p)~\cos(\psi_p/2)$}, where $v_p$ is the magnitude of the tridimensional velocity vector, $\gamma_p$ is the aspect angle referenced to the bistatic bisector, and $\psi_p$ is the bistatic angle, all associated with the point target \cite{skolnik2008}. To ensure that targets are correctly illuminated, the beamforming directions at \ac{gNB} \#1 and \ac{gNB} \#2 can be previously estimated via hierarchical beam search \cite{li2021_search} or other possibly more efficient strategies such as an extension of the sampling and reconstruction of the angular domain proposed in \cite{mandelli2023} to the bistatic case. Furthermore, power allocation  among the dedicated beams to the reference path or to the sensing of targets can be performed aiming to meet communication and radar sensing performance requirements \cite{zhang2019,liyanaarachchi2021_allocation}.
	
	At \ac{gNB} \#1, a discrete-frequency domain transmit frame \mbox{$\mathbf{X}\in\mathbb{C}^{N\times M}$} with \mbox{$M\in\mathbb{N}_{>0}$} \ac{OFDM} symbols, each with \mbox{$N\in\mathbb{N}_{>0}$} subcarriers, is generated. Each \ac{OFDM} symbol undergoes \ac{IDFT} and has \ac{CP} of length \mbox{$N_\mathrm{CP}\in\mathbb{N}_{\geq0}$} prepended to it. The resulting discrete-time domain sequence \mbox{$x[s]\in\mathbb{C}$}, \mbox{$s\in\{0,1,\dots,M(N+N_\mathrm{CP})-1\}$}, from \ac{P/S} conversion on $\mathbf{X}$ then undergoes \ac{D/A} conversion with sampling rate $F_\mathrm{s}^\mathrm{Tx}$, whose corresponding sampling period is \mbox{$T_\mathrm{s}^\mathrm{Tx}=1/F_\mathrm{s}^\mathrm{Tx}$}. This results in a baseband transmit signal \mbox{$x(t)\in\mathbb{C}$} that occupies a bandwidth \mbox{$B\leq F_\mathrm{s}^\mathrm{Tx}$}. Consequently, the subcarrier spacing in the frequency domain is \mbox{$\Delta f=B/N$}. After conditioning by the \ac{AFE}, transmission at the carrier frequency \mbox{$f_\text{c}\gg B$}, and reception at \ac{gNB} \#2, the baseband receive signal \mbox{$y(t)\in\mathbb{C}$} at the receiving \ac{ISAC} node in the considered system is expressed as in \eqref{eq:yt}. In this equation, the attenuation factor $\alpha_0$, which accounts for all losses in the reference path from \ac{gNB} \#1 to \ac{gNB} \#2, is given by
	\begin{figure*}[!b]
		\hrulefill
		\vspace*{4pt}
		\setcounter{equation}{3}
		\begin{equation}\label{eq:woSFO_freq}
		\tilde{Y}_{n,m} \approx~\alpha_0~X_{n,m}\e^{\im\psi_\Delta}+\sum_{p=1}^{P-1}\alpha_p~X_{n,m}~\e^{-\im 2\pi n\Delta f(\tau_p-\tau_0)}~\e^{\im 2\pi f_{\mathrm{D},p}[m(N+N_\mathrm{CP})+N_\mathrm{CP}]/B}~\e^{\im\psi_\Delta}
		\end{equation}
		\hrulefill
		\vspace*{4pt}
		\setcounter{equation}{10}
		\begin{equation}\label{eq:SFO_freq_xi}
		\xi^\mathrm{SFO,ICI}_{n,m} =~\sum_{l=-N/2,l\neq n}^{N/2-1}\tilde{Y}_{l,m}~\frac{\sin\left(\pi\left[\left(1+\delta\right)l-n\right]\right)}{N\sin\left(\frac{\pi\left[\left(1+\delta\right)l-n\right]}{N}\right)}~\e^{\im 2\pi\frac{m\left(N+N_\mathrm{CP}\right)+N_\mathrm{CP}}{N}\delta l}~\e^{\im\pi\frac{N-1}{N}\left[\left(1+\delta\right)l-n\right]}
		\end{equation}
	\end{figure*}
	\setcounter{equation}{1}
	\begin{equation}\label{eq:alpha_c}
	\alpha_0 = \sqrt{\frac{G_\mathrm{Tx}G_\mathrm{Rx}\lambda_0^2}{\left(4\pi\right)^3R_0^4}},
	\end{equation}
	
	\noindent where $G_\mathrm{Tx}$ and $G_\mathrm{Rx}$ are the combined single-antenna and beamforming gains at \acp{gNB} \#1 and \#2, respectively.  In addition, \mbox{$\lambda_0=c_0/f_\text{c}$} is the wavelength associated with the carrier frequency $f_\text{c}$. The attenuation factor $\alpha_p$ for the $p\mathrm{th}$ propagation path associated with a point target can be expressed as
	\setcounter{equation}{2} 
	\begin{equation}\label{eq:alpha_s}
	\alpha_p = \sqrt{\frac{G_\mathrm{Tx}G_\mathrm{Rx}\sigma_{\mathrm{RCS},p}\lambda_0^2}{\left(4\pi\right)^3{R^\text{T---Rx}_p}^2{R^\text{T---Rx}_p}^2}}.
	\end{equation}
	In this equation, $\sigma_{\mathrm{RCS},p}$ denotes the bistatic \ac{RCS} of the point target associated with the $p\mathrm{th}$ path, which is aspect-dependent. In other words, $\sigma_{\mathrm{RCS},p}$ depends on the angle from which the target is illuminated by \ac{gNB} \#1 and the angle to which reflections are directed towards \ac{gNB} \#2. Regarding $\tau_\Delta$, $f_\Delta$, and $\psi_\Delta$, they respectively represent the \ac{STO} $\tau_{\Delta}$ caused by the mismatch between transmitter and receiver time references, the \ac{CFO} $f_{\Delta}$ and its resulting phase rotation $\psi_\Delta$ between the oscillators at transmitter and receiver \ac{ISAC} nodes.
	
	Next, it is assumed that the baseband signal $y(t)$ is sampled with the sampling frequency \mbox{$F_\mathrm{s}^\mathrm{Rx}$} at the receiver at the time instants \mbox{$sT_\mathrm{s}^\mathrm{Rx}$}, where \mbox{$T_\mathrm{s}^\mathrm{Rx}=1/F_\mathrm{s}^\mathrm{Rx}$} is the sampling period at the receiver. The aforementioned sampling produces the sequence \mbox{$y[s]\in\mathbb{C}$}. 
	For the sake of simplicity, it is assumed that the adopted sampling rate $F_\mathrm{s}^\mathrm{Rx}$ at the receiver of the bistatic \ac{ISAC} system is exactly the same as $F_\mathrm{s}^\mathrm{Tx}$ adopted at the transmitter side, which results in the non-occurrence of \ac{SFO}. It is further assumed that the length $N_\mathrm{CP}$ of the \ac{CP} prepended to the \ac{OFDM} symbols at the transmitter side is sufficient to avoid \ac{ISI} due to propagation, and that \ac{STO} and \ac{CFO} are estimated and corrected. The aforementioned synchronization can be performed, e.g., using the reference path and relying on fully known preamble \ac{OFDM} symbols \cite{schmidl1997} or, as in the \ac{5G} \ac{NR} case, partly known preamble \ac{OFDM} symbols \cite{omri2019}. Consequently, a receive \ac{OFDM} frame \mbox{$\tilde{\mathbf{Y}}\in\mathbb{C}^{N\times M}$} can be formed by removing \ac{CP} and performing a \ac{DFT} on each \ac{OFDM} symbol extracted from the resulting samples after performing time and frequency synchronization on $y[s]$. The element $\tilde{Y}_{n,m}\in\mathbb{C}$ at the $n\mathrm{th}$ row, \mbox{$n\in\{-N/2,1,\dots,N/2-1\}$}, and $m\mathrm{th}$ column, \mbox{$m\in\{0,1,\dots,M-1\}$}, of $\tilde{\mathbf{Y}}$ can be expressed as in \eqref{eq:woSFO_freq} \cite{giroto2023_EuMW}. In this equation, $X_{n,m}\in\mathbb{C}$ is the element at the $n\mathrm{th}$ row and $m\mathrm{th}$ column of the transmit \ac{OFDM} frame in the discrete-frequency domain $\mathbf{X}$.
	In practice, however, \ac{SFO} occurs, which leads to different $F_\mathrm{s}^\mathrm{Rx}$ and $F_\mathrm{s}^\mathrm{Tx}$ and, consequently, to degradation of the receive \ac{OFDM} frame $\tilde{\mathbf{Y}}$. Consequently, the \ac{SFO} must be estimated and corrected before communication processing to estimate the transmit frame $\mathbf{X}$ and subsequent bistatic radar sensing processing as described in \cite{giroto2023_EuMW} can be performed.
	
	
	\section{Pilot-Based SFO Estimation and Correction}\label{sec:SFO_sec}
	
	If the adopted sampling rate $F_\mathrm{s}^\mathrm{Rx}$ to perform analog-to-digital conversion on $y(t)$ to produce $y[s]$ at the receiver side is different than $F_\mathrm{s}^\mathrm{Tx}$, then \ac{SFO} occurs. The experienced \ac{SFO} is given by
	\setcounter{equation}{4}
	\begin{equation}
	\mathrm{SFO} = F_\mathrm{s}^\mathrm{Rx}-F_\mathrm{s}^\mathrm{Tx},
	\end{equation}
	and normalized \ac{SFO} by the transmit sampling rate $F_\mathrm{s}^\mathrm{Tx}$ is defined as $\delta\in\mathbb{R}$ and expressed as
	\begin{equation}\label{eq:delta}
	\delta = \frac{\mathrm{SFO}}{F_\mathrm{s}^\mathrm{Tx}} = \frac{F_\mathrm{s}^\mathrm{Rx}-F_\mathrm{s}^\mathrm{Tx}}{F_\mathrm{s}^\mathrm{Tx}}.
	\end{equation}
	For a non-zero normalized \ac{SFO} $\delta$, $y[s]$ becomes
	\begin{equation}\label{eq:rx_sig_discrete_SFO}
	y[s] = y(t)\Big|_{t=sT_\mathrm{s}^\mathrm{Tx}(1-\delta)} + w[s],
	\end{equation}
	where \mbox{$w[s]\in\mathbb{C}$} is the sampled \ac{AWGN}. Consequently, the receive \ac{OFDM} frame $\tilde{\mathbf{Y}}$ still assuming fully corrected \ac{STO} and \ac{CFO} becomes \mbox{$\mathbf{Y}\in\mathbb{C}^{N\times M}$}. For small \acp{SFO} that do lead to \ac{ISI}, the element \mbox{$Y_{n,m}\in\mathbb{C}$} at the $n\mathrm{th}$ row and $m\mathrm{th}$ column of $\mathbf{Y}$ can be expressed as \cite{bookSFO,chiueh2012}
	\begin{equation}\label{eq:SFO_freq}
	Y_{n,m} = \alpha^\mathrm{SFO}_{n}\left(\tilde{Y}_{n,m}\right)\e^{\im\psi^\mathrm{SFO}_{n,m}} + \xi^\mathrm{SFO,ICI}_{n,m} + W_{n,m},
	\end{equation}
	where \mbox{$\alpha^\mathrm{SFO}_{n}\in\mathbb{R}_{\geq 0}$} is an amplitude modulation expressed as
	\begin{equation}\label{eq:SFO_freq_alpha}
	\alpha^\mathrm{SFO}_{n} = \frac{\sin\left(\pi\delta n\right)}{N\sin\left(\frac{\pi\delta n}{N}\right)},
	\end{equation}
	\mbox{$\psi^\mathrm{SFO}_{n,m}\in\mathbb{R}$} is a phase rotation expressed as
	\begin{equation}\label{eq:SFO_freq_psi}
	\psi^\mathrm{SFO}_{n,m} = -\frac{2\pi\delta n}{N}\left[m\left(N+N_\mathrm{CP}\right)+N_\mathrm{CP}\right]-\frac{\pi\delta n}{N}(N-1),
	\end{equation}
	and \mbox{$\xi^\mathrm{SFO,ICI}_{n,m}\in\mathbb{C}$} is the \ac{ICI} term expressed as in \eqref{eq:SFO_freq_xi}, and \mbox{$W_{n,m}\in\mathbb{C}$} is the equivalent sampled \ac{AWGN} at the $n\mathrm{th}$ subcarrier of the $m\mathrm{th}$ \ac{OFDM} symbol in the frame.
	
	Rearranging the first term of \eqref{eq:SFO_freq_psi} and resorting to the \ac{DFT} shift theorem, it is concluded that the \ac{SFO} causes a delay that is equal to
	\setcounter{equation}{11}
	\begin{equation}\label{eq:SFO_tau_m}
	\tau^\mathrm{SFO}_{m} = \delta\left[m(N+N_\mathrm{CP})+N_\mathrm{CP}\right]T_\mathrm{s}
	\end{equation}
	at the $m\mathrm{th}$ \ac{OFDM} symbol. Since $\tau^\mathrm{SFO}_{m}$ changes with $m$, a delay migration defined as
	\begin{equation}\label{eq:SFO_delta_tau_m}
	\Delta\tau^\mathrm{SFO}_{m} = \tau^\mathrm{SFO}_{m} - \tau^\mathrm{SFO}_{0} = \delta\left[m(N+N_\mathrm{CP})\right]T_\mathrm{s}
	\end{equation}
	is experienced along the \ac{OFDM} symbols in $\mathbf{Y}$. Comparing \eqref{eq:SFO_freq} with the effect of a \ac{CFO} on the receive \ac{OFDM} symbol as described in \cite{bookSFO}, it is also concluded that the normalized \ac{SFO} $\delta$ causes a frequency shift equal to
	\begin{equation}\label{eq:SFO_f_n}
	f^\mathrm{SFO}_{n} = (\delta n)\Delta f
	\end{equation}
	at the $n\mathrm{th}$ subcarrier of every \ac{OFDM} symbol, which corresponds to the $n\mathrm{th}$ row of $\mathbf{Y}$. As $f^\mathrm{SFO}_{n}$ is a function of $n$, a frequency shift migration
	\begin{equation}\label{eq:SFO_freq_mig}
	\Delta f^\mathrm{SFO}=f^\mathrm{SFO}_{N/2-1}-f^\mathrm{SFO}_{-N/2}
	\end{equation}
	is experienced from the first to the last \ac{OFDM} subcarrier in each \ac{OFDM} symbol contained in $\mathbf{Y}$.

	By estimating the previously described \ac{SFO}-induced impairments to the reference path, the \ac{SFO} itself can be estimated. For that purpose, either preamble \ac{OFDM} symbols \cite{giroto2023_EuMW,brunner2024} or pilot subcarriers can be used under the condition that they are know at the receiver side. The first approach, however, leads to undesirable overhead, reducing the communication data rate. By using only pilot subcarriers, a possible approach to estimate the \ac{SFO} is by observing the frequency shift migration along consecutive at pilot subcarriers along the \ac{OFDM} symbols. To improve the effective \ac{SNR} for \ac{SFO} estimation, the delay migration can observed instead. This approach consists of obtained \ac{CFR} estimates from the the pilot subcarriers in $\mathbf{Y}$ \cite{hsieh1998,coleri2002}, and transforming them into \ac{CIR} estimates via \acp{IDFT}. From these \ac{CIR} estimates, the delay migration, and consequently, the \ac{SFO} can be estimated. This approach benefits from a processing gain that is to the one experienced during range processing in \ac{OFDM}-based radar and \ac{ISAC} systems \cite{giroto2021_tmtt}. Consequently, higher effective \ac{SNR} and better accuracy for \ac{SFO} estimation is experienced than in the approach based on frequency shift migration that operates on the subcarrier level. 
	
	An \ac{SFO} estimation approach based on delay migration has been proposed by Wu et al.~\cite{wu2012}, which however has not provided accuracy bounds to their estimator. In this sense, Section~\ref{subsec:SFO_delay_est} describes an \ac{SFO} estimation approach based on \cite{wu2012}, highlighting its limitations and providing accuracy bounds. To ensure higher robustness at \ac{SFO}-induced \ac{ISI} and reduce computational complexity at higher \acp{SFO}, the \ac{TITO} method is presented in Section~\ref{subsec:SFO_TITO_est}. Finally, Section~\ref{subsec:SFO_corr} discusses \ac{SFO} correction strategies that can be performed once it has been estimated. 
		
	\subsection{Delay migration-based SFO estimation}\label{subsec:SFO_delay_est}
	
	\begin{figure}[!t]
		\centering
		\resizebox{8cm}{!}{
			\psfrag{A}[c][c]{\footnotesize $0$}
			\psfrag{B}[c][c]{\footnotesize $1$}
			\psfrag{H}[c][c]{\footnotesize $\Delta N_\mathrm{pil}-1$}
			\psfrag{I}[c][c]{\footnotesize $\Delta N_\mathrm{pil}$}
			\psfrag{J}[c][c]{\footnotesize $\Delta N_\mathrm{pil}+1$}
			\psfrag{K}[c][c]{\footnotesize $N-1$}
			
			\psfrag{EE}[c][c]{\footnotesize $0$}
			\psfrag{FF}[c][c]{\footnotesize $1$}
			\psfrag{GG}[c][c]{\footnotesize $2$}
			\psfrag{M-3}[c][c]{\footnotesize $\Delta M_\mathrm{pil}-1$}
			\psfrag{M-2}[c][c]{\footnotesize $\Delta M_\mathrm{pil}$}
			\psfrag{M-1}[c][c]{\footnotesize $\Delta M_\mathrm{pil}+1$}
			\psfrag{N}[c][c]{\footnotesize $M-1$}
			
			\psfrag{WWW}[c][c]{\footnotesize $\Delta N_\mathrm{pil}$}
			\psfrag{ZZZ}[c][c]{\footnotesize $\Delta M_\mathrm{pil}$}
			
			\psfrag{x}[c][c]{}
			\psfrag{y}[c][c]{}
			
			\includegraphics[width=7.5cm]{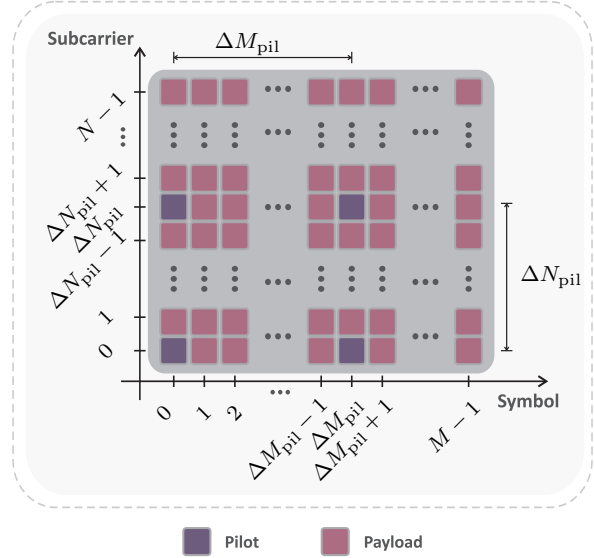}
		}
		\captionsetup{justification=raggedright,labelsep=period,singlelinecheck=false}
		\caption{\ Structure of the OFDM frame $\mathbf{X}$ in the discrete-frequency domain.}\label{fig:frameStructure}
		
	\end{figure}
	
	\begin{figure*}[!b]
		\hrulefill
		\vspace*{4pt}
		\setcounter{equation}{19}
		\begin{equation}\label{eq:SFO_LS_1}
			\widehat{\delta} = \argmin_{\delta} \sum_{m_\mathrm{pil}=0}^{M_\mathrm{pil}-1} \left(\widehat{\Delta\tau}^\mathrm{SFO}_{\mathrm{pil},m_\mathrm{pil}} - \left\{\delta\left[(m_\mathrm{pil}\Delta M_\mathrm{pil})(N+N_\mathrm{CP})\right]T_\mathrm{s}\right\}\right)^2
		\end{equation}
		\vspace*{4pt}
		\hrulefill
		\vspace*{4pt}
		\setcounter{equation}{20}
		\begin{equation}\label{eq:SFO_LS_2}
			\widehat{\delta} = \frac{M_\mathrm{pil}\sum_{m_\mathrm{pil}=0}^{M_\mathrm{pil}-1} \left[(m_\mathrm{pil}\Delta M_\mathrm{pil})(N+N_\mathrm{CP})T_\mathrm{s}\right] \widehat{\Delta\tau}^\mathrm{SFO}_{\mathrm{pil},m_\mathrm{pil}} - \sum_{m_\mathrm{pil}=0}^{M_\mathrm{pil}-1} \left[(m_\mathrm{pil}\Delta M_\mathrm{pil})(N+N_\mathrm{CP})T_\mathrm{s}\right] \sum_{m_\mathrm{pil}=0}^{M_\mathrm{pil}-1} \widehat{\Delta\tau}^\mathrm{SFO}_{\mathrm{pil},m_\mathrm{pil}}}{M_\mathrm{pil}\sum_{m_\mathrm{pil}=0}^{n}\left[(m_\mathrm{pil}\Delta M_\mathrm{pil})(N+N_\mathrm{CP})T_\mathrm{s}\right]^2 - \left[\sum_{m_\mathrm{pil}=0}^{M_\mathrm{pil}-1} (m_\mathrm{pil}\Delta M_\mathrm{pil})(N+N_\mathrm{CP})T_\mathrm{s}\right]^2}
		\end{equation}
		\vspace*{4pt}
		\hrulefill
		\vspace*{4pt}
		\setcounter{equation}{21}
		\begin{equation}\label{eq:SFO_LS_3}
			\sigma_{\widehat{\delta}} \approx \frac{\sigma_{\widehat{\Delta\tau}^\mathrm{SFO}_{\mathrm{pil},m_\mathrm{pil}}}}{\sqrt{M_\mathrm{pil}\sum_{m_\mathrm{pil}=0}^{n}\left[(m_\mathrm{pil}\Delta M_\mathrm{pil})(N+N_\mathrm{CP})T_\mathrm{s}\right]^2 - \left[\sum_{m_\mathrm{pil}=0}^{M_\mathrm{pil}-1} (m_\mathrm{pil}\Delta M_\mathrm{pil})(N+N_\mathrm{CP})T_\mathrm{s}\right]^2}}
		\end{equation}
		\vspace*{4pt}
		\hrulefill
		\vspace*{4pt}
		\setcounter{equation}{24}
		\begin{equation}\label{eq:SFO_LS_4}
		\sigma_{\widehat{\delta}} \gtrapprox \sqrt{\frac{\{6/[\mathrm{SNR}(N^2-1)N]\}[1/(2\pi\Delta f\Delta N_\mathrm{pil})]^2}{M_\mathrm{pil}\sum_{m_\mathrm{pil}=0}^{n}\left[(m_\mathrm{pil}\Delta M_\mathrm{pil})(N+N_\mathrm{CP})T_\mathrm{s}\right]^2 - \left[\sum_{m_\mathrm{pil}=0}^{M_\mathrm{pil}-1} (m_\mathrm{pil}\Delta M_\mathrm{pil})(N+N_\mathrm{CP})T_\mathrm{s}\right]^2}}
		\end{equation}
		\vspace*{4pt}
		\hrulefill
		\vspace*{4pt}
		\setcounter{equation}{26}
		\begin{equation}\label{eq:sqrt_mle_tau_2}
			\sigma_{\widehat{\delta}} \gtrapprox \frac{\sqrt{3}/(6\eta N_\mathrm{pil}\Delta f\Delta N_\mathrm{pil})}{\sqrt{M_\mathrm{pil}\sum_{m_\mathrm{pil}=0}^{n}\left[(m_\mathrm{pil}\Delta M_\mathrm{pil})(N+N_\mathrm{CP})T_\mathrm{s}\right]^2 - \left[\sum_{m_\mathrm{pil}=0}^{M_\mathrm{pil}-1} (m_\mathrm{pil}\Delta M_\mathrm{pil})(N+N_\mathrm{CP})T_\mathrm{s}\right]^2}}
		\end{equation}
	\end{figure*}
	
	Assuming sufficient \ac{SNR} regime, negligible \ac{SFO}-induced \ac{ICI} and no \ac{SFO}-induced \ac{ISI} at every $m\mathrm{th}$ \ac{OFDM} symbol, one can use the pilot subcarriers in $\mathbf{Y}$ to obtain \ac{CIR} estimates. In this article, they are assumed to be uniformly distributed for simplicity, as depicted in Fig.~\ref{fig:frameStructure}. In practical implementations such as \ac{5G} \ac{NR}, \ac{DMRS} \cite{kanhere2021_multistatic} could be used as the aforementioned pilot subcarriers at first assessment, as it is present in sufficient density in the \ac{OFDM} frame and does not introduce communication overhead. However, the beamforming coefficients associated with \ac{DMRS} defined at \ac{gNB} \#1 are different across the subcarriers and not known at the receiver in \ac{gNB} \#2 due to \ac{UE}-specific precoding. Consequently, this may prevent using \ac{DMRS} to obtain the required \ac{CIR} estimates for \ac{SFO} estimation. An alternative solution is to repurpose \ac{PRS} \cite{dwivedi2021,wei2023} subcarriers for \ac{SFO} estimation, as it adopts known, constant beamforming coefficients.

	 The obtained \ac{CIR} estimates from the pilot subcarriers in $\mathbf{Y}$ form a delay-slow time profile represented by the matrix \mbox{$\mathbf{D}\in\mathbb{C}^{N_\mathrm{pil}\times M_\mathrm{pil}}$}. To ensure fine delay granularity in the contained \ac{CIR} estimates in $\mathbf{D}$ and therefore improve the accuracy of estimated delays, the required \ac{IDFT} processing can be zero padded. The \mbox{$m_\mathrm{pil}\mathrm{th}$} column, \mbox{$m_\mathrm{pil}\in\{0,1,\dots,M_\mathrm{pil}-1\}$}, of the $\mathbf{D}$ contains one \ac{CIR} estimate obtained from the $N_\mathrm{pil}$ pilot subcarriers at the corresponding pilot \ac{OFDM} symbol. Assuming spacing $\Delta N_\mathrm{pil}$ and $\Delta M_\mathrm{pil}$ between pilot subcarriers in the column and row directions in the \ac{OFDM} frame, it holds that
	\setcounter{equation}{15}
	\begin{equation}
	N_\mathrm{pil} = \left\lfloor\frac{N}{\Delta N_\mathrm{pil}}\right\rfloor+1,
	\end{equation}
	and
	\begin{equation}
	M_\mathrm{pil} = \left\lfloor\frac{M}{\Delta M_\mathrm{pil}}\right\rfloor+1.
	\end{equation}
	Since the assumed uniform pilot subcarrier spacing may not be observed in practical deployments, it is also worth highlighting that it is not required for delay migration and consequent \ac{SFO} estimation.
	
	To estimate the residual \ac{SFO}, the proposed method, \ac{TITO}, is applied to the reference path enabled by the relay \ac{RIS}. In this context, the aforementioned path is taken as a benchmark and its delay progression along the succesive \ac{CIR} estimates contained in the columns of $\mathbf{D}$ is analyzed. For that purpose, the delay migration of the aforementioned path is estimated at every $m_\mathrm{pil}\mathrm{th}$ column of $\mathbf{D}$, yielding $\widehat{\Delta\tau}^\mathrm{SFO}_{\mathrm{pil},m_\mathrm{pil}}$. Knowing that the correspondence between column indexes from $\mathbf{D}$ and $\mathbf{Y}$ is given by
	\begin{equation}
	m = m_\mathrm{pil}\Delta M_\mathrm{pil},
	\end{equation}
	it holds that
	\begin{equation}
	\widehat{\Delta\tau}^\mathrm{SFO}_{\mathrm{pil},m_\mathrm{pil}} = \widehat{\Delta\tau}^\mathrm{SFO}_{m_\mathrm{pil}\Delta M_\mathrm{pil}}.
	\end{equation}
	In this equation, $\widehat{\Delta\tau}^\mathrm{SFO}_{m_\mathrm{pil}\Delta M_\mathrm{pil}}$ denotes an estimate of $\Delta\tau^\mathrm{SFO}_{m_\mathrm{pil}\Delta M_\mathrm{pil}}$. By feeding all $M_\mathrm{pil}$ delay migration estimates $\widehat{\Delta\tau}^\mathrm{SFO}_{\mathrm{pil},m_\mathrm{pil}}$ to a linear regression model based on \eqref{eq:SFO_delta_tau_m}, a \ac{LS} estimate $\widehat{\delta}$ of the normalized \ac{SFO} is finally obtained. This is achieved by taking the partial derivative of the sum of the squared differences to be minimized in \eqref{eq:SFO_LS_1} w.r.t. $\delta$, setting it to zero and solving to yield $\widehat{\delta}$, which is expressed as in \eqref{eq:SFO_LS_2}.

	Due to the linear regression nature of the \ac{SFO} estimation, the estimate $\widehat{\delta}$ has a standard deviation $\sigma_{\widehat{\delta}}$ given by \eqref{eq:SFO_LS_3}. This equation shows that $\sigma_{\widehat{\delta}}$ depends on the standard deviation $\sigma_{\widehat{\Delta\tau}^\mathrm{SFO}_{\mathrm{pil},m_\mathrm{pil}}}$ of the delay migration estimates at each $m_\mathrm{pil}\mathrm{th}$ column of $\mathbf{D}$. Based on the \ac{CRLB} of delay migration estimation in \ac{OFDM}-based systems at high \ac{SNR} \cite{braun2011}, the lower bound for $\sigma_{\widehat{\Delta\tau}^\mathrm{SFO}_{\mathrm{pil},m_\mathrm{pil}}}$ assuming negligible \ac{SFO}-induced \ac{ICI} and \ac{ISI} is
	\setcounter{equation}{22}
	\begin{equation}\label{eq:sqrt_crlb_tau}
	\sigma_{\widehat{\Delta\tau}^\mathrm{SFO}_{\mathrm{pil},m_\mathrm{pil}}} \geq \sqrt{\frac{6}{\mathrm{SNR}(N_\mathrm{pil}^2-1)N_\mathrm{pil}}\left(\frac{1}{2\pi\Delta f \Delta N_\mathrm{pil}}\right)^2}.
	\end{equation}
	The \ac{SNR} in this equation is calculated assuming that the receive signal in the considered \ac{RIS}-assisted bistatic \ac{ISAC} system is impaired by \ac{AWGN}, being therefore given by 
	\begin{equation}\label{eq:SNR}
	\mathrm{SNR} = \frac{P_\mathrm{Tx}\alpha_\mathrm{c}^2\beta_\mathrm{c}^2}{k_\mathrm{B}~B~T_\mathrm{therm}~\mathrm{NF}},
	\end{equation}
	where  $k_\text{B}$ is the Boltzmann constant, $T_\text{therm}$ is the standard room temperature in Kelvin, and $\mathit{NF}$ is the overall receiver noise figure. 
	Based on \eqref{eq:SFO_LS_3} and \eqref{eq:sqrt_crlb_tau}, a lower bound for $\sigma_{\widehat{\delta}}$ at high \ac{SNR} and still assuming negligible \ac{SFO}-induced \ac{ICI} and \ac{ISI} can be defined as in \eqref{eq:SFO_LS_4}. While this lower bound can only be reached by unbiased estimators such as \ac{MUSIC}. In practical deployments, however, if the processing based on zero-padded \acp{IDFT} is used as a \ac{MLE}, then bias occurs due to the limited granularity of the axis of possible delays that can be estimated: This axis is solely composed by integer multiples of \mbox{$1/(\eta B)$}, where \mbox{$\eta\in\mathbb{N}_{\geq1}$} is the \ac{ZP} factor. Consequently, a more accurate lower bound for \mbox{$\sigma_{\widehat{\Delta\tau}^\mathrm{SFO}_{\mathrm{pil},m_\mathrm{pil}}}$} at high \ac{SNR} is given by \cite{braun2011}
	\setcounter{equation}{25}
	\begin{equation}\label{eq:sqrt_mle_tau}
	\sigma_{\widehat{\Delta\tau}^\mathrm{SFO}_{\mathrm{pil},m_\mathrm{pil}}} \geq \frac{\sqrt{3}}{6\eta N_\mathrm{pil}\Delta f\Delta N_\mathrm{pil}}.
	\end{equation}
	Consequently, a lower bound for $\sigma_{\widehat{\delta}}$ for the case in which $\widehat{\delta}$ is obtained based on \ac{MLE} delay migration estimates $\widehat{\Delta\tau}^\mathrm{SFO}_{\mathrm{pil},m_\mathrm{pil}}$ can be defined as in \eqref{eq:sqrt_mle_tau_2}. Since the \ac{MLE} error is actually uniformly distributed due to the quantized zero-padded \ac{IDFT} bins, the aforementioned lower bound is only a simplified approximation.
	
	In order for the \ac{SFO} estimation to work as intended, i.e., without relevant performance degradation or bias, the experienced normalized \ac{SFO} $\delta$ should not induce \ac{ICI} and \ac{ISI} that result in relevant degradation of the reference path in the \acp{CIR} estimates in the columns of $\mathbf{D}$. Some remarks on these issues are presented as follows.
	
	\begin{thm_rem}
		It is ideally expected that normalized \ac{SFO} $\delta$ only results in frequency shifts $f^\mathrm{SFO}_{n}$ that cause negligible \ac{ICI}.
	\end{thm_rem}
	\begin{proof}
		To avoid non-negligible \ac{ICI} that will eventually lead to degradation of the delay and consequently \ac{SFO} estimation and its subsequent correction, the \ac{SFO}-induced frequency shift $f^\mathrm{SFO}_{n}$ at every $n\mathrm{th}$ \ac{OFDM} subcarriers must be significantly lower than the subcarrier spacing $\Delta f$. Assuming the typical frequency shift upperbound for \ac{OFDM} systems of one tenth of $\Delta f$ \cite{hwang2009,giroto2021_tmtt},
		\setcounter{equation}{27}
		\begin{equation}\label{eq:ub_fshift}
		f^\mathrm{SFO}_{n} < \Delta f/10
		\end{equation}
		should hold to result in negligible \ac{ICI}. Combining \eqref{eq:ub_fshift} with \eqref{eq:SFO_f_n} results in $(\delta n)\Delta f < \Delta f/10$. Applying \eqref{eq:ub_fshift} to the highest absolute \ac{SFO}-induced \ac{ICI}, which is experienced at $n=\pm N/2$, results in
		\begin{equation}\label{eq:maxSFO_ICI}
		\lvert\delta\rvert<\delta^\mathrm{ICI}_\mathrm{max},
		\end{equation}
		where
		\begin{equation}
		\delta^\mathrm{ICI}_\mathrm{max} = 1/(5N).
		\end{equation}
		Note that, while subcarrier $n=-N/2$ exists, $n=N/2$ is only approached when performing zero-padding.
	\end{proof}
	
	\begin{thm_rem}
		It is ideally expected that the normalized \ac{SFO} $\delta$ only results in delays $\tau^\mathrm{SFO}_{m}$ that do not cause \ac{ISI}.
	\end{thm_rem}
	\begin{proof}
		To avoid \ac{ISI}-induced performance degradation of the delay and \ac{SFO} estimation and therefore of the \ac{RIS}-assisted bistatic \ac{ISAC} system as a whole, \ac{ISI} should not happen at any of the \ac{OFDM} symbols in the frame. If the \ac{SFO}-induced delay $\tau^\mathrm{SFO}_{m}$ is negative, \ac{ISI} will necessarily occur. For positive delays, however, $\tau^\mathrm{SFO}_{m}$ at all \ac{OFDM} symbols, regardless of whether they contain pilots or not, should be lower than the \ac{CP}-free delay $\tau^\mathrm{CP}_\mathrm{max}=N_\mathrm{CP}/B$. Therefore, it should hold that
		\begin{equation}
		\tau^\mathrm{CP}_\mathrm{max} > \tau^\mathrm{SFO}_{m_\mathrm{pil}\Delta M_\mathrm{pil}} \geq 0.
		\end{equation}
		Consequently, the delay $\tau^\mathrm{SFO}_{M-1}$ at the $(M-1)\mathrm{th}$ column of $\mathbf{Y}$ should be such that
		\begin{equation}
		N_\mathrm{CP}/B > \tau^\mathrm{SFO}_{M-1} \geq 0 .
		\end{equation}
		Based on \eqref{eq:SFO_tau_m}, this results in
		\begin{equation}\label{eq:maxSFO_ISI}
		\delta^\mathrm{ISI}_\mathrm{max} > \delta \geq 0 ,
		\end{equation}
		where
		\begin{equation}
		\delta^\mathrm{ISI}_\mathrm{max} = \frac{N_\mathrm{CP}}{B\left[(M-1)(N+N_\mathrm{CP})+N_\mathrm{CP}\right]T_\mathrm{s}}.
		\end{equation}
	\end{proof}
	
	%
	
	\subsection{TITO method for ISI-aware SFO estimation}\label{subsec:SFO_TITO_est}
	
	As discussed in Section~\ref{subsec:SFO_delay_est}, the presented approximate lower bounds for the \ac{SFO} estimates in \eqref{eq:SFO_LS_4} and \eqref{eq:sqrt_mle_tau_2} so far are only accurate under the assumption of negligible \ac{SFO}-induced \ac{ICI} and \ac{ISI}. The aforementioned \ac{ICI} is expected to be handled by the previously mentioned processing gain of $\SI[parse-numbers = false]{10\log_{10}(N_\mathrm{pil})}{dB}$ at the delay migration estimation. The \ac{SFO}-induced \ac{ISI}, however, can severely degrade the delay migration estimation \cite{wang2023} and, consequently, the \ac{SFO} estimation. In order to avoid this, the \ac{TITO} method for \ac{SFO} estimation is proposed. This method consists of performing the described \ac{SFO} estimation based on the slope (or tilt) estimation (or inference) of the delay (or time offset) migration of the reference path along \ac{CIR} estimates contained in the columns of $\mathbf{D}$ in Section~\ref{subsec:SFO_delay_est}, which ultimately allows obtaining an estimate $\widehat{\delta}$ of the normalized \ac{SFO} based on the knowledge of its relationship with the delay migration described by \eqref{eq:SFO_delta_tau_m}. The novelty of \ac{TITO} compared to apporaches such as the one based on the method by Wu et al.~\cite{wu2012} lies in the adaptive choice of the number $M_\mathrm{pil}^\mathrm{TITO}\leq M_\mathrm{pil}$ of columns of $\mathbf{D}$ to be used for delay migration and subsequent \ac{SFO} estimation with the aim of avoiding using \ac{CIR} estimates the originate from \ac{ISI}-contamined pilot \ac{OFDM} symbols, which is explained as follows. First, it is important to highlight that \ac{ISI} only happens at the pilot \ac{OFDM} symbols of index $m_\mathrm{pil}$ where the \ac{SFO}-induced delay $\tau^\mathrm{SFO}_{m_\mathrm{pil}}$ is either negative or positive and exceeds the \ac{CP} duration $N_\mathrm{CP}T_\mathrm{s}^\mathrm{Tx}$. In both cases, the \ac{ISI}-induced \ac{SINR} degradation of the reference path in the aforementioned \ac{CIR} estimates depends on the \ac{SFO}-induced excess delay causing the \ac{ISI}. Considering these issues, only the $M_\mathrm{pil}^\mathrm{TITO}$ first columns of $\mathbf{D}$ where tolerable \ac{SINR} degradation is experienced are used to obtain the required delay migration estimates for the \ac{SFO} estimation with the \ac{TITO} method. This also results in lower computational complexity compared with the method by Wu et al.~\cite{wu2012} in cases with high \ac{SFO}, since only a limited\ac number of columns of $\mathbf{D}$ are used for \ac{CIR} and subsequent \ac{SFO} estimation. Based on \eqref{eq:SFO_delta_tau_m}, $M_\mathrm{pil}^\mathrm{TITO}$ is chosen such that
	\resizebox{\columnwidth}{!}{
		\begin{minipage}{\columnwidth}
			\begin{align}\label{eq:M_TITO}
			M_\mathrm{pil}^\mathrm{TITO} =& \max \quad m_\mathrm{pil}+1 \nonumber\\
			&\text{s.t.} \quad \left\lvert \frac{\widehat{\Delta\tau}^\mathrm{SFO}_{\mathrm{pil},m_\mathrm{pil}}-\widehat{\Delta\tau}^\mathrm{SFO}_{\mathrm{pil},m_\mathrm{pil}-1}}{\Delta M_\mathrm{pil}(N+N_\mathrm{CP})T_\mathrm{s}^\mathrm{Tx}} \right\rvert \leq (1+\varepsilon)\lvert\delta_\mathrm{max}\rvert.\nonumber\\
			\end{align}
		\end{minipage}
	}
	\vspace{0.1cm}
	In \eqref{eq:M_TITO}, \mbox{$\lvert\delta_\mathrm{max}\rvert\in\mathbb{R}_{\geq 0}$} is the maximum absolute normalized \ac{SFO} (positive or negative) that is expected in the \ac{RIS}-assisted bistatic \ac{ISAC} system. Furthermore, $\varepsilon$ denotes an empirically defined margin that accounts for factors that influence the delay migration estimation inaccuracies, such as the chosen \ac{ZP} factor $\eta$ in the \ac{MLE} estimation and the \ac{SNR}. If the constraint in \eqref{eq:M_TITO} is violated, it is considered that the \ac{SFO}-induced \ac{ISI} causes significant \ac{SINR} degradation that in turn leads to non-negligible bias in the delay migration estimates at pilot \ac{OFDM} symbols of index $m_\mathrm{pil}=M_\mathrm{pil}^\mathrm{TITO}$ or higher. The \acp{CIR} contained in $\mathbf{D}$ which correspond to these pilot \ac{OFDM} symbols are therefore discarded and only the $M_\mathrm{pil}^\mathrm{TITO}$ first columns of $\mathbf{D}$ are considered for the delay migration estimation and subsequent \ac{SFO} estimation. While this degrades the accuracy of the \ac{SFO} estimation according to \eqref{eq:SFO_LS_4} and \eqref{eq:sqrt_mle_tau_2}, it avoids biasing the \ac{SFO} estimate $\widehat{\delta}$ due to \ac{ISI} as later shown in the results in Section~\ref{sec:simResults}. Finally, it is worth highlighting that, forcing \mbox{$M_\mathrm{pil}^\mathrm{TITO}=M_\mathrm{pil}$}, the \ac{TITO} \ac{SFO} estimation method becomes equivalent to approach based on the method by Wu et al.~\cite{wu2012}.
		
	\subsection{SFO correction}\label{subsec:SFO_corr}
	
	Once the estimate $\widehat{\delta}$ has been obtained with \eqref{eq:SFO_LS_2} using a limited number of pilot \ac{OFDM} symbols defined according to \eqref{eq:M_TITO}, resampling is performed on the discrete-time domain samples associated with the receive frame including \ac{CP}. Typically, a Farrow filter \cite{farrow1988} with previous cubic interpolation \cite{erup1993} is employed. After resampling, \ac{CP} is removed and the \ac{OFDM} symbols are once again transformed into the discrete-frequency domain for further communication processing to estimate the transmit frame $\mathbf{X}$ and subsequent bistatic radar sensing processing as described in \cite{giroto2023_EuMW}.
	
	While the resampling approach is able to correct both low and high \acp{SFO}, a simplified correction can be performed for the case where the \ac{SFO} is so low that negligible \ac{ICI} and \ac{ISI} occur. In this case, a \ac{ZF} equalization of $Y_{n,m}$ is performed to yield an estimate of $\tilde{Y}_{n,m}$. This estimate is given by \mbox{$\widehat{\tilde{Y}}_{n,m} = Y_{n,m}/(\widehat{\alpha}^\mathrm{SFO}_{n}\e^{\im\widehat{\psi}^\mathrm{SFO}_{n,m}})$}, where $\widehat{\alpha}^\mathrm{SFO}_{n}$ and $\widehat{\psi}^\mathrm{SFO}_{n,m}$ are estimates of $\alpha^\mathrm{SFO}_{n}$ and $\psi^\mathrm{SFO}_{n,m}$ calculated by using $\widehat{\delta}$ instead of its true value $\delta$ in \eqref{eq:SFO_freq_alpha} and \eqref{eq:SFO_freq_psi}, respectively. Since this method is assumed to be used for very low \ac{SFO} only, \mbox{$\alpha^\mathrm{SFO}_{n}\approx 1$} is expected for all $n$, and the described equalization can be simplified to \mbox{$\widehat{\tilde{Y}}_{n,m} = Y_{n,m}/\e^{\im\widehat{\psi}^\mathrm{SFO}_{n,m}}$} or, alternatively,
	\begin{align}\label{eq:SFO_freq_corr}
	\widehat{\tilde{Y}}_{n,m} \approx \tilde{Y}_{n,m} + W_{n,m}/\e^{\im\widehat{\psi}^\mathrm{SFO}_{n,m}}.
	\end{align}
	The $N$ rows and $M$ columns filled with elements $\widehat{\tilde{Y}}_{n,m}$ form a receive frame \mbox{$\mathbf{\widehat{\tilde{Y}}}\in\mathbb{C}^{N\times M}$}, which can finally be used for the previously mentioned further processing. Since this approach only requires dividing each subcarrier in the discrete-frequency domain by a phase term, it results in lower computational complexity than imposed by resampling approach. This happens since the latter requires resampling the whole sequence $y[s]$, including samples belonging to the \acp{CP} of the \ac{OFDM} symbols. In addition, \ac{DFT} must be performed again for each \ac{OFDM} symbol after serial-to-parallel conversion and \ac{CP} removal to reconstruct the \ac{OFDM} frame with corrected \ac{SFO}.
	
	%
	
	
	\begin{table}[!t]
		\renewcommand{\arraystretch}{1.5}
		\arrayrulecolor[HTML]{708090}
		\setlength{\arrayrulewidth}{.1mm}
		\setlength{\tabcolsep}{4pt}
		
		\centering
		\captionsetup{width=43pc,justification=centering,labelsep=newline}
		\caption{\textsc{Adopted OFDM Signal Parameters}}
		\label{tab:ofdmParameters}
			\begin{tabular}{|cc|}
				\hhline{|==|}
				\multicolumn{1}{|c|}{\textbf{Carrier frequency} ($f_\text{c}$)}      & $\SI{26.2}{\giga\hertz}$ \\ \hline
				\multicolumn{1}{|c|}{\textbf{Frequency bandwidth} ($B$)}      & $\SI{500}{\mega\hertz}$ \\ \hline
				\multicolumn{1}{|c|}{\textbf{No. of subcarriers} ($N$)}      & $2048$ \\ \hline
				\multicolumn{1}{|c|}{\textbf{Subcarrier spacing} ($\Delta f$)}      & $\SI{244.14}{\kilo\hertz}$ \\ \hline
				\multicolumn{1}{|c|}{\textbf{CP length} ($N_\mathrm{CP}$)}      & $512$ \\ \hline
				\multicolumn{1}{|c|}{\textbf{No. of OFDM symbols} ($M$)}      & $4096$ \\ \hline
				\multicolumn{1}{|c|}{\textbf{Pilot spacing} ($\Delta N_\mathrm{pil}$, $\Delta M_\mathrm{pil}$)}      & $2$, $4$ \\ \hline
				\multicolumn{1}{|c|}{\textbf{No. of pilot subc. per pilot OFDM symbol} ($N_\mathrm{pil}$)}      & $1024$ \\ \hline
				\multicolumn{1}{|c|}{\textbf{No. of OFDM symbols w/ pilots} ($M_\mathrm{pil}$)}      & $1024$ \\  \hline
				\multicolumn{1}{|c|}{\textbf{Digital modulation}}      & QPSK \\  \hline
				\multicolumn{1}{|c|}{\textbf{Channel coding and code rate}}      & LDPC, $2/3$ \\ \hhline{|==|}	
				
			\end{tabular}
		\vspace{-0.10cm}
	\end{table}
					
	\section{Numerical Results}\label{sec:simResults}
	
	In this section, the performance of the \ac{RIS}-assisted bistatic \ac{ISAC} system with the introduced \ac{TITO} \ac{SFO} estimation method is analyzed. For this purpose, the \ac{OFDM} signal parameters listed in Table~\ref{tab:ofdmParameters}, which according to \cite{giroto2021_tmtt,giroto2023_EuMW} result in the radar and communication performance parameters listed in Table~\ref{tab:performanceParameters}, are adopted. While these parameters do not comply, e.g., with \ac{5G} \ac{NR} numerology, they were chosen to meet constraints of the measurement setup described later in Section~\ref{sec:measResults}, allowing to achieve suitable communication and radar sensing performances.
	
	\begin{table}[!t]
		\renewcommand{\arraystretch}{1.5}
		\arrayrulecolor[HTML]{708090}
		\setlength{\arrayrulewidth}{.1mm}
		\setlength{\tabcolsep}{4pt}
		
		\centering
		\captionsetup{width=43pc,justification=centering,labelsep=newline}
		\caption{\textsc{Communication and Radar Performance Parameters}}
		\label{tab:performanceParameters}
		\begin{tabular}{|cc|}
			\hhline{|==|}
			\multicolumn{1}{|c|}{\textbf{Comm. data rate} (100\% duty cycle, $\mathcal{R}_\mathrm{comm}$)}      & $\SI{0.40}{Gbit/s}$ \\  \hline
			\multicolumn{1}{|c|}{\textbf{Processing gain} ($G_\mathrm{p}$)}  & $\SI{69.24}{dB}$ \\ \hline
			\multicolumn{1}{|c|}{\textbf{Range resolution} ($\Delta R$)}     & $\SI{0.60}{\meter}$ \\ \hline
			\multicolumn{1}{|c|}{\textbf{Max. unamb. range} ($R_\mathrm{max,ua}$)}  & $\SI{1227.95}{\meter}$ \\ \hline
			\multicolumn{1}{|c|}{\textbf{Max. ISI-free range} ($R_\mathrm{max,ISI}$)}    & $\SI{306.99}{\meter}$ \\ \hline
			\multicolumn{1}{|c|}{\textbf{Doppler shift resolution} ($\Delta f_\mathrm{D}$)}  & $\SI{47.68}{\hertz}$ \\ \hline
			\multicolumn{1}{|c|}{\textbf{Max. unamb. Doppler shift} ($f_\mathrm{D,max,ua}$)}  & $\SI{97.66}{\kilo\hertz}$ \\ \hline
			\multicolumn{1}{|c|}{\textbf{Max. ICI-free Doppler shift} ($f_\mathrm{D,max,ICI}$)} & $\SI{24.41}{\kilo\hertz}$ \\ \hhline{|==|}		
			
		\end{tabular}
	\end{table}
	
	\begin{figure}[!t]
		\centering
		\psfrag{AAAA}[c][c]{\scriptsize $10^{-1}$}
		\psfrag{BBBB}[c][c]{\scriptsize $10^{0}$}
		\psfrag{CCCC}[c][c]{\scriptsize $10^{1}$}
		\psfrag{DDDD}[c][c]{\scriptsize $10^{2}$}
		\psfrag{EEEE}[c][c]{\scriptsize $10^{3}$}
		
		\psfrag{FFFF}[c][c]{\scriptsize $10^{-2}$}
		\psfrag{GGGG}[c][c]{\scriptsize $10^{-1}$}
		\psfrag{HHHH}[c][c]{\scriptsize $10^{0}$}
		\psfrag{IIII}[c][c]{\scriptsize $10^{1}$}
		\psfrag{JJJJ}[c][c]{\scriptsize $10^{2}$}
		\psfrag{KKKK}[c][c]{\scriptsize $10^{3}$}
		
		\psfrag{XXXXXXXXXX}{\small $\lvert \delta\rvert$ (ppm)}
		\psfrag{YYYYYYYYYYYY}{\small $\lvert \mathrm{SFO}\rvert$ (kHz)}
		
		\includegraphics[width=5.5cm]{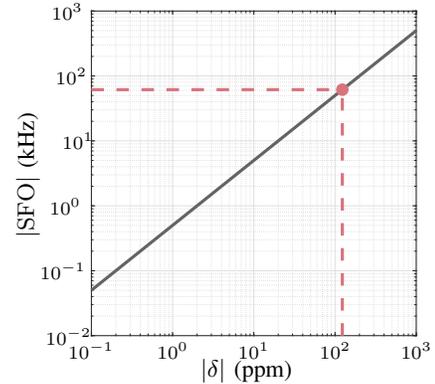}
		
		\captionsetup{justification=raggedright,labelsep=period,singlelinecheck=false}
		\caption{\ Equivalent SFO in kHz to the normalized SFO to the Nyquist sampling rate $F_\mathrm{s}=B$ ({\color[rgb]{0.3922,0.3922,0.3922}\textbf{\textemdash}}). For reference, the maximum expected absolute value of the SFO in the bistatic setup from Section~\ref{sec:measResults} based on two Zynq UltraScale+ RFSoC ZCU111 \ac{SoC} platforms  is also shown ({\color[rgb]{0.8471,0.4510,0.4980}$\CIRCLE$}).}\label{fig:sfo_ppm_kHz}
		\vspace{-0.30cm}
	\end{figure}

	\begin{figure*}[!t]
		\centering
		\subfloat[ ]{
			\psfrag{-1024}[c][c]{\scriptsize -$1024$}
			\psfrag{-512}[c][c]{\scriptsize -$512$}
			\psfrag{0}[c][c]{\scriptsize $0$}
			\psfrag{512}[c][c]{\scriptsize $512$}
			\psfrag{768}[c][c]{\scriptsize $768$}
			\psfrag{1023}[c][c]{\scriptsize $1023$}
			
			\psfrag{-0.1}[c][c]{\scriptsize -$0.1$}
			\psfrag{-0.2}[c][c]{\scriptsize -$0.2$}
			
			\psfrag{0}[c][c]{\scriptsize $0$}
			\psfrag{-10}[c][c]{\scriptsize -$10$}
			\psfrag{-20}[c][c]{\scriptsize -$20$}
			\psfrag{-30}[c][c]{\scriptsize -$30$}
			
			\psfrag{XX}{$n$}
			\psfrag{YYYYYYYYYYYYY}{$\alpha^\mathrm{SFO}_{n}$ (dB)}
			
			\includegraphics[height=4.5cm]{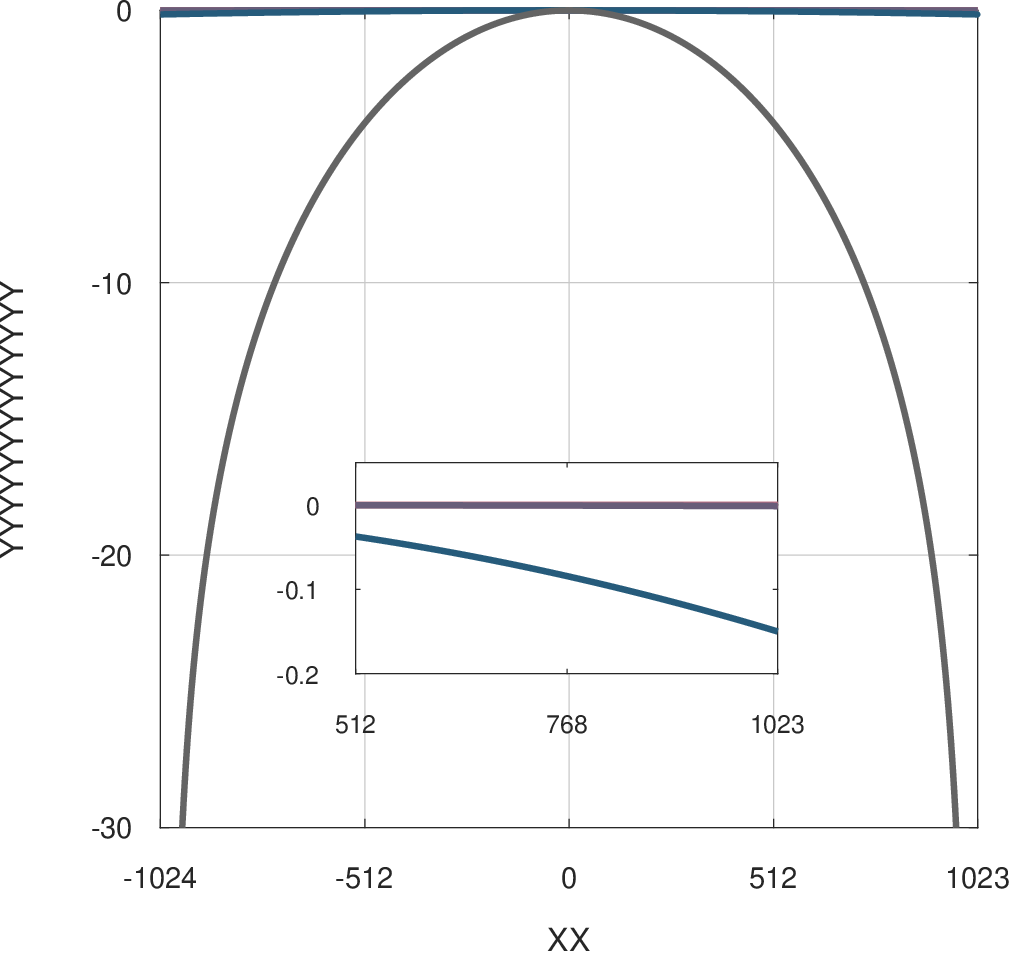}
		}\hspace{0.25cm}
		\subfloat[ ]{
			\psfrag{0}[c][c]{\scriptsize $0$}
			\psfrag{1023}[c][c]{\scriptsize $1023$}
			\psfrag{2047}[c][c]{\scriptsize $2047$}
			\psfrag{3071}[c][c]{\scriptsize $3071$}
			\psfrag{4095}[c][c]{\scriptsize $4095$}
			
			\psfrag{GGGG}[c][c]{\scriptsize $10^{-6}$}
			\psfrag{HHHH}[c][c]{\scriptsize $10^{-4}$}
			\psfrag{IIII}[c][c]{\scriptsize $10^{-2}$}
			\psfrag{JJJJ}[c][c]{\scriptsize $10^{0}$}
			\psfrag{KKKK}[c][c]{\scriptsize $10^{2}$}
			
			\psfrag{LLLL}[c][c]{\scriptsize $3\times10^{-4}$}
			\psfrag{MMMM}[c][c]{\scriptsize $3\times10^{-2}$}
			\psfrag{NNNN}[c][c]{\scriptsize $3$}
			\psfrag{OOOO}[c][c]{\scriptsize $3\times10^{2}$}
			\psfrag{PPPP}[c][c]{\scriptsize $3\times10^{4}$}
			
			\psfrag{XX}{$m$}
			\psfrag{YYYYYYYY}{\hspace{-0.4cm}$\lvert\Delta\tau^\mathrm{SFO}_{m}\rvert$ ($\mu$s)}
			\psfrag{ZZZZZZZZZZ (m)}{\hspace{-0.3cm}$\lvert c_0\Delta\tau^\mathrm{SFO}_{m}\rvert$ (m)}
			
			\includegraphics[height=4.5cm]{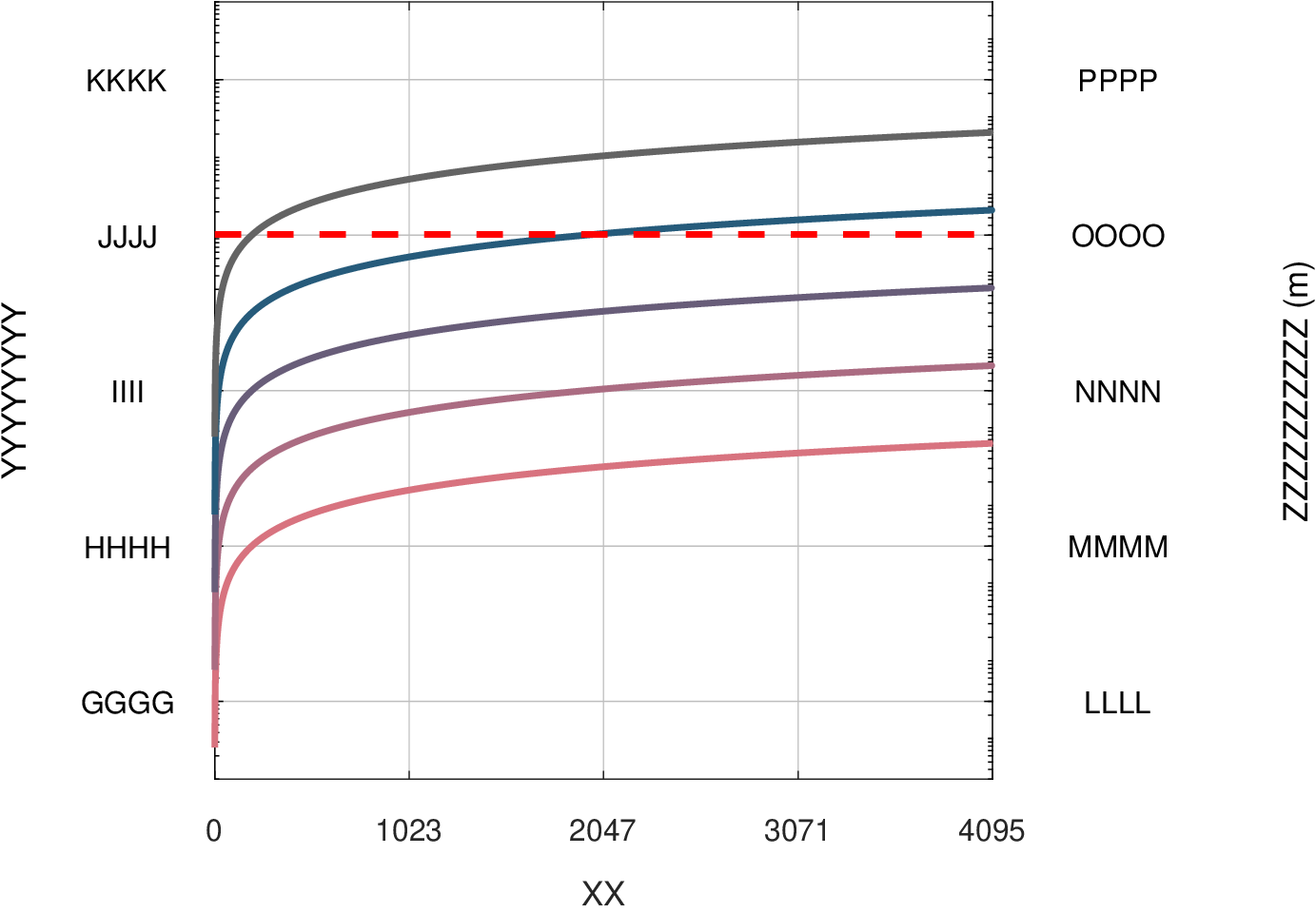}
		}\hfill
		\subfloat[ ]{
			\psfrag{-1024}[c][c]{\scriptsize -$1024$}
			\psfrag{-512}[c][c]{\scriptsize -$512$}
			\psfrag{0}[c][c]{\scriptsize $0$}
			\psfrag{512}[c][c]{\scriptsize $512$}
			\psfrag{768}[c][c]{\scriptsize $768$}
			\psfrag{1023}[c][c]{\scriptsize $1023$}
			
			\psfrag{GGGG}[c][c]{\scriptsize $10^{-5}$}
			\psfrag{HHHH}[c][c]{\scriptsize $10^{-3}$}
			\psfrag{IIII}[c][c]{\scriptsize $10^{-1}$}
			\psfrag{JJJJ}[c][c]{\scriptsize $10^{1}$}
			\psfrag{KKKK}[c][c]{\scriptsize $10^{3}$}
			
			\psfrag{XX}{$n$}
			\psfrag{YYYYYYYY}{\hspace{-0.35cm}$\lvert f_n\rvert$ (kHz)}
			
			\includegraphics[height=4.5cm]{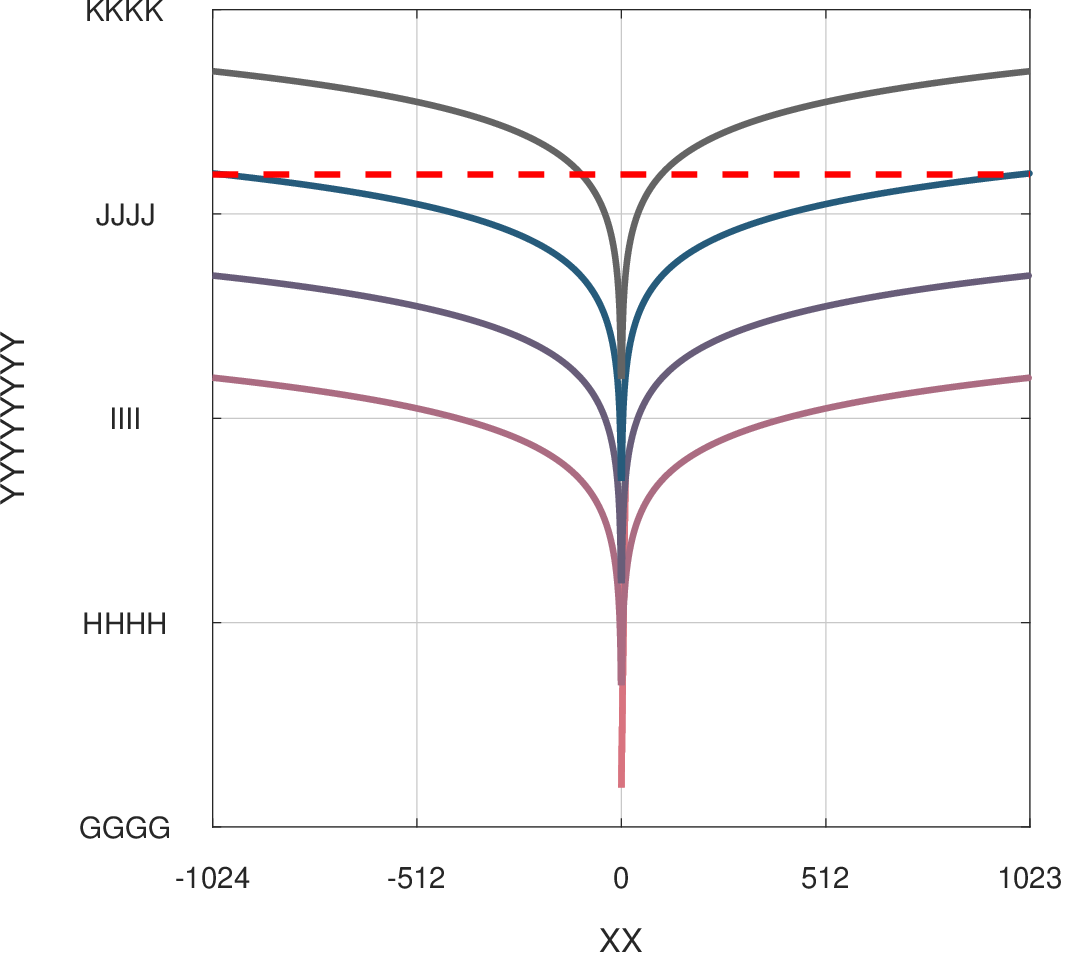}
		}
		
		\captionsetup{justification=raggedright,labelsep=period,singlelinecheck=false}
		\caption{\ SFO effects: (a) amplitude modulation $\alpha^\mathrm{SFO}_{n}$ along the subcarriers, (b) absolute value of the delay migration $\Delta\tau^\mathrm{SFO}_{m}$ along the \ac{OFDM} symbols, and (c) absolute value of the subcarrier frequency shift $f^\mathrm{SFO}_{n}$ along the subcarriers, all evaluated for $\lvert \delta\rvert=\SI{0.1}{ppm}$ ({\color[rgb]{0.8471,0.4510,0.4980}\textbf{\textemdash}}), $\lvert \delta\rvert=\SI{1}{ppm}$ ({\color[rgb]{0.6706,0.4235,0.5098}\textbf{\textemdash}}), $\lvert \delta\rvert=\SI{10}{ppm}$ ({\color[rgb]{0.4078,0.3647,0.4745}\textbf{\textemdash}}), $\lvert \delta\rvert=\SI{100}{ppm}$ ({\color[rgb]{0.1490,0.3569,0.4824}\textbf{\textemdash}}), and $\lvert \delta\rvert=\SI{1000}{ppm}$ ({\color[rgb]{0.5294,0.5294,0.5294}\textbf{\textemdash}}) For comparison with the achieved values, the maximum \ac{ISI}-free delay of $N_\mathrm{CP}/B$ and range of $c_0N_\mathrm{CP}/B$ for the positive $\delta$ case, as well as the maximum tolerable frequency shift of $\Delta f/10$ defined in \eqref{eq:ub_fshift} which causes negligible \ac{ICI} are marked ({\color[rgb]{0.8471,0.4510,0.4980}\textbf{\textendash~\textendash}}) in (b) and (c), respectively.}\label{fig:SFO_limitations}
		
	\end{figure*}
	
	\begin{figure}[!t]
		\centering
		
		\psfrag{-1.5}[c][c]{\scriptsize -$1.5$}
		\psfrag{-1}[c][c]{\scriptsize -$1$}
		\psfrag{-0.5}[c][c]{\scriptsize -$0.5$}
		\psfrag{0}[c][c]{\scriptsize $0$}
		\psfrag{0.5}[c][c]{\scriptsize $0.5$}
		\psfrag{1}[c][c]{\scriptsize $1$}
		\psfrag{1.5}[c][c]{\scriptsize $1.5$}
		
		\psfrag{I}{\small $I$}
		\psfrag{Q}{\small $Q$}
		
		\includegraphics[width=5.5cm]{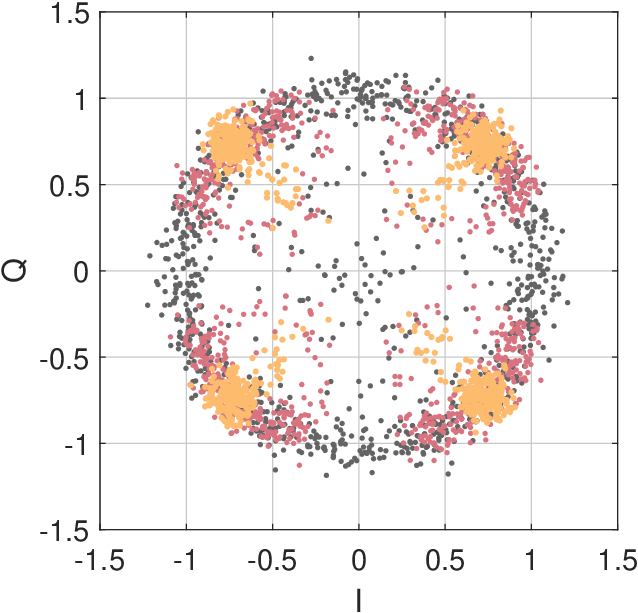}

		\captionsetup{justification=raggedright,labelsep=period,singlelinecheck=false}
		
		\caption{\ Normalized receive \ac{QPSK} constellations obtained from simulations for an \ac{AWGN} channel with $\mathrm{SNR} = \SI{20}{dB}$ and $\delta=\SI{1}{ppm}$ at \ac{OFDM} symbols $m=0$ ({\color[rgb]{0.9882,0.7333,0.4275}$\CIRCLE$}, \mbox{$\mathrm{EVM}=\SI{-21.39}{dB}$}), $m=64$ ({\color[rgb]{0.8471,0.4510,0.4980}$\CIRCLE$}, \mbox{$\mathrm{EVM}=\SI{-13.05}{dB}$}), and $m=4094$ ({\color[rgb]{0.3922,0.3922,0.3922}$\CIRCLE$}, \mbox{$\mathrm{EVM}=\SI{0.18}{dB}$}).}\label{fig:QPSK_const_sim}
		
	\end{figure}
		
	\subsection{SFO-induced impairment of ISAC system performance}\label{subsec:sfoImpConst}
	
	Based on the adopted parameters, the corresponding \ac{SFO} in kHz to $\delta$, which represents the normalized \ac{SFO} to the Nyquist sampling rate $F_\mathrm{s}=B$, is shown in Fig.~\ref{fig:sfo_ppm_kHz}. In this figure, the maximum normalized \ac{SFO} of \mbox{$\pm\SI{123}{ppm}$} expected with the adopted measurement setup later described in Section~\ref{sec:measResults} based on two distinct Zynq UltraScale+ RFSoC ZCU111 \ac{SoC} platforms is also shown. This number derives from the stability of \mbox{$\pm\SI{61.5}{ppm}$} of the programmable clocks at each board, from which the sampling clocks are derived, and it represents a common normalized \ac{SFO} expected in practice.
	
	Next, the \ac{SFO}-induced impairments to the discrete-frequency domain \ac{OFDM} frame $\mathbf{Y}$ apart from \ac{ISI} discussed in Section~\ref{sec:SFO_sec}, namely the amplitude modulation $\alpha^\mathrm{SFO}_{n}$, the delay migration $\Delta\tau^\mathrm{SFO}_{m}$, and the subcarrier frequency shift $f^\mathrm{SFO}_{n}$, are shown in Fig~\ref{fig:SFO_limitations} for \mbox{$\lvert \delta\rvert\in\SI[parse-numbers = false]{\{0, 0.1, 1, 10, 100, 1000\}}{ppm}$}. It can be seen in Fig.~\ref{fig:SFO_limitations}(a) that all considered \ac{SFO} values except for \mbox{$\lvert \delta\rvert=\SI{1000}{ppm}$} yield negligible amplitude modulation along the subcarrier index $n$. In Figs.~\ref{fig:SFO_limitations}(b) and (c), however, it is seen that \mbox{$\lvert \delta\rvert\in\SI[parse-numbers = false]{\{100,1000\}}{ppm}$} yield delays and frequency shifts that exceed both the \ac{ICI}- and \ac{ISI}-free \ac{SFO} limits defined according to \eqref{eq:maxSFO_ICI} and \eqref{eq:maxSFO_ISI}, respectively. Based on the \ac{OFDM} signal parameters listed in Table~\ref{tab:ofdmParameters}, the aforementioned \ac{SFO} ranges for approximately \ac{ICI}- and \ac{ISI}-free operation are $\SI{97.66}{ppm}>\delta>\SI{-97.66}{ppm}$ and $\SI{48.83}{ppm}>\delta>\SI{0}{ppm}$, respectively.	
	The achieved results show that only $\lvert \delta\rvert\in\SI[parse-numbers = false]{\{100,1000\}}{ppm}$ among the considered normalized \ac{SFO} values exceed both \ac{ICI}- and \ac{ISI}-free \ac{SFO} ranges due to their resulting delays and frequency shifts, respectively, as previously discussed and shown in Figs.~\ref{fig:SFO_limitations}(b) and (c). Since the achieved \ac{ICI}- and \ac{ISI}-free \ac{SFO} ranges do not fully cover the expected \acp{SFO} of up to \mbox{$\pm\SI{123}{ppm}$} that may occur in the aforementioned measurement setup that will later be considered in Section~\ref{sec:measResults}, the \ac{SFO} correction via the frequency-domain equalization from \eqref{eq:SFO_freq_corr} is henceforth not considered and the resampling approach explained in Section~\ref{subsec:SFO_corr} is adopted instead.
	
	\begin{figure}[!t]
		\centering
		
		\psfrag{55}{(a)}
		\psfrag{22}{(b)}
		
		\psfrag{0}[c][c]{\scriptsize $0$}
		\psfrag{4}[c][c]{\scriptsize $4$}
		\psfrag{8}[c][c]{\scriptsize $8$}
		\psfrag{12}[c][c]{\scriptsize $12$}	
		
		\psfrag{-10}[c][c]{\scriptsize -$10$}
		\psfrag{-5}[c][c]{\scriptsize -$5$}
		\psfrag{0}[c][c]{\scriptsize $0$}
		\psfrag{5}[c][c]{\scriptsize $5$}
		\psfrag{10}[c][c]{\scriptsize $10$}
		
		\psfrag{0}[c][c]{\scriptsize $0$}
		\psfrag{-15}[c][c]{\scriptsize -$15$}
		\psfrag{-30}[c][c]{\scriptsize -$30$}
		\psfrag{-45}[c][c]{\scriptsize -$45$}
		\psfrag{-60}[c][c]{\scriptsize -$60$}
		
		\psfrag{Doppler shift (kHz)}[c][c]{Doppler shift (kHz)}
		\psfrag{Rel. bist. range (m)}[c][c]{\small Rel. bist. range (m)}
		\psfrag{Norm. mag. (dB)}[c][c]{\small Norm. mag. (dB)}
		
		\includegraphics[width=8.75cm]{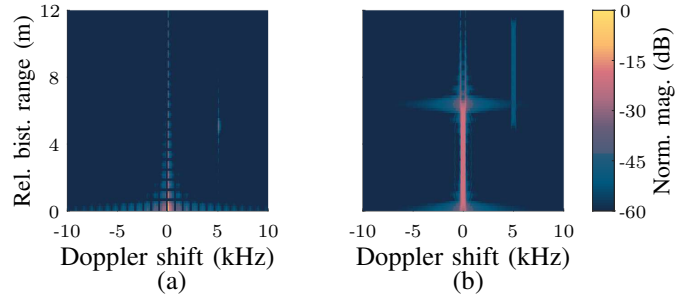}
		\captionsetup{justification=raggedright,labelsep=period,singlelinecheck=false}
		\caption{\ Range-Doppler bistatic radar images with one reference path and a target at a relative bistatic range of $\SI{5}{\meter}$ and Doppler shift of $\SI{5}{\kilo\hertz}$ for (a) $\delta=\SI{0}{ppm}$ and (b) $\delta=\SI{1}{ppm}$.}\label{fig:I_full}
		
	\end{figure}
	\begin{figure}[!t]
		\centering
		
		\psfrag{55}{(a)}
		\psfrag{22}{(b)}
		
		\psfrag{0}[c][c]{\scriptsize $0$}
		\psfrag{2}[c][c]{\scriptsize $2$}
		\psfrag{4}[c][c]{\scriptsize $4$}
		\psfrag{6}[c][c]{\scriptsize $6$}
		\psfrag{8}[c][c]{\scriptsize $8$}
		
		\psfrag{-0.5}[c][c]{\scriptsize -$0.5$}
		\psfrag{-0.25}[c][c]{\scriptsize -$0.25$}
		\psfrag{0}[c][c]{\scriptsize $0$}
		\psfrag{0.25}[c][c]{\scriptsize $0.25$}
		\psfrag{0.5}[c][c]{\scriptsize $0.5$}
		
		\psfrag{0}[c][c]{\scriptsize $0$}
		\psfrag{-15}[c][c]{\scriptsize -$15$}
		\psfrag{-30}[c][c]{\scriptsize -$30$}
		\psfrag{-45}[c][c]{\scriptsize -$45$}
		\psfrag{-60}[c][c]{\scriptsize -$60$}
		
		\psfrag{Doppler shift (kHz)}[c][c]{Doppler shift (kHz)}
		\psfrag{Rel. bist. range (m)}[c][c]{\small Rel. bist. range (m)}
		\psfrag{Norm. mag. (dB)}[c][c]{\small Norm. mag. (dB)}
		
		\includegraphics[width=8.75cm]{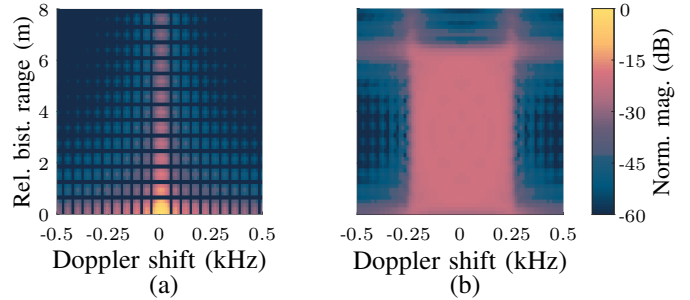}
		\captionsetup{justification=raggedright,labelsep=period,singlelinecheck=false}
		\caption{\ Zoom-in on reference path of the radar images from Fig.~\ref{fig:I_full} for (a) $\delta=\SI{0}{ppm}$ and (b) $\delta=\SI{1}{ppm}$.}\label{fig:I_zoom}
	\end{figure}
	
	To clearly illustrate the effects of \ac{SFO} on the receive \ac{QPSK} constellation, simulation results for an \ac{AWGN} channel with $\mathrm{SNR}=\SI{20}{dB}$ and an \ac{SFO} of $\delta=\SI{1}{ppm}$ are shown in Fig.~\ref{fig:QPSK_const_sim}. The assumed channel model agrees with the assumption that the reference path between transmitter and receiver is much stronger than the propagation paths associated with radar targets and that perfect time and frequency synchronization w.r.t. this main path is performed. In this aforementioned figure, the \ac{ICI} effect described in \eqref{eq:SFO_freq_xi} and shown in Fig.~\ref{fig:SFO_limitations}(c) can be observed for the \ac{OFDM} symbol of index $m=0$ in the form of spreading of some constellation points towards the origin, while the constellation for $m=64$ is still clearly affected by \ac{ICI} but also presents more relevant phase rotation due to the dependence of $\psi^\mathrm{SFO}_{n,m}$ in \eqref{eq:SFO_freq_psi} on the \ac{OFDM} symbol index $m$. As for \ac{OFDM} symbol $m=4094$, both effects are present, but the higher \ac{SFO}-induced delay results in more accentuated rotation of the \ac{QPSK} constellation.

	Regarding sensing performance, bistatic radar images assuming $\mathrm{SNR}=\SI{20}{dB}$ before radar processing and perfect time and frequency synchronization w.r.t. the reference path, as well as a radar target with $\mathrm{SNR}=\SI{-10}{dB}$ before radar processing, relative range of $\SI{5}{\meter}$ and relative Doppler shift of $\SI{5}{\kilo\hertz}$ w.r.t. the main path are shown in Fig.~\ref{fig:I_full} for $\delta=\SI{0}{ppm}$ and $\delta=\SI{1}{ppm}$. In both range and Doppler shift directions, rectangular windowing was used to avoid that the \ac{SFO} effects are masked by the use of a window aiming to reduce sidelobe level. For simplicity, the presented radar images are obtained by processing all subcarriers. In practice, this would require pilot-based channel estimation and equalization to correctly estimate the \ac{QPSK} data in payload subcarriers and only then allow bistatic radar signal processing with the full frame as described in \cite{giroto2023_EuMW,brunner2024}. The results in these figures show that range and Doppler shift migration occur for $\delta=\SI{1}{ppm}$, spreading both the reference path and the radar target in the relative bistatic range and Doppler shift directions and reducing their magnitude and consequently \ac{SINR} compared with the case with $\delta=\SI{0}{ppm}$. These effects are respectively caused by delay and frequency shift migrations described by \eqref{eq:SFO_delta_tau_m} and \eqref{eq:SFO_freq_mig}. A zoom in on the reference path for both cases is shown in Fig.~\ref{fig:I_zoom}. In this figure, a range migration from $\SI{0}{\meter}$ to \mbox{$c_0\{\delta[M(N+N_\mathrm{CP})]T_\mathrm{s}\}=\SI{6.29}{\meter}$} is observed, which agrees with the expected delay migration that can be calculated based on \eqref{eq:SFO_delta_tau_m}. Similarly, a Doppler shift migration from \mbox{$\delta(-N/2)\Delta f=\SI{-0.25}{\kilo\hertz}$} to \mbox{$\delta(N/2)\Delta f=\SI{0.25}{\kilo\hertz}$} is experienced as predicted in \eqref{eq:SFO_f_n} and \eqref{eq:SFO_freq_mig}. Besides range and Doppler shift migration, both the reference path and the radar target experience \ac{SINR} loss due to the \ac{SFO}-induced \ac{ICI}. \ac{ISI}, however, is not experienced since $\delta=\SI{1}{ppm}$ lies in the \ac{ISI}-free \ac{SFO} range $\SI{48.83}{ppm}>\delta>\SI{0}{ppm}$ defined by \eqref{eq:maxSFO_ISI} for the \ac{OFDM} parameterization in the considered \ac{RIS}-assisted bistatic \ac{ISAC} system.
			
	The presented \ac{QPSK} constellations and radar images for $\delta=\SI{1}{ppm}$ show that even small \acp{SFO} can greatly impair both communication and radar sensing performances of bistatic \ac{ISAC} systems. For the expected \ac{SFO} of up to $\delta=\pm\SI{123}{ppm}$ in the measurement setup later described in Section~\ref{sec:measResults}, even higher degradation is expected since \ac{ISI} can occur even for positive \acp{SFO} as indicated by Fig.~\ref{fig:SFO_limitations}(b). To prevent this, the experienced \ac{SFO} must be accurately estimated and corrected, which is addressed in Section~\ref{subsec:sfoEst}.
	
	\begin{figure}[!t]
		\centering
		
		\psfrag{-30}[c][c]{\scriptsize $-30$}
		\psfrag{-20}[c][c]{\scriptsize $-20$}
		\psfrag{-10}[c][c]{\scriptsize $-10$}
		\psfrag{0}[c][c]{\scriptsize $0$}
		\psfrag{10}[c][c]{\scriptsize $10$}
		\psfrag{20}[c][c]{\scriptsize $20$}
		\psfrag{30}[c][c]{\scriptsize $30$}
		
		\psfrag{AAAA}[c][c]{\scriptsize $10^{-4}$}
		\psfrag{BBBB}[c][c]{\scriptsize $10^{-2}$}
		\psfrag{CCCC}[c][c]{\scriptsize $10^{0}$}
		\psfrag{DDDD}[c][c]{\scriptsize $10^{2}$}
		\psfrag{EEEE}[c][c]{\scriptsize $10^{4}$}
		
		\psfrag{XXXXXXX}{\small SNR (dB)}
		\psfrag{YYYYYYYYYYYY}{\small RMSE (ppm)}
		
		\includegraphics[width=5.5cm]{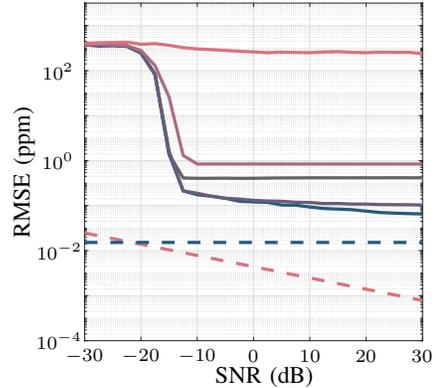}
		
		\captionsetup{justification=raggedright,labelsep=period,singlelinecheck=false}
		\caption{\ RMSE of delay-based SFO estimation based on \cite{wu2012} for $\mathrm{SNR} = \SI{-30}{dB}$ through $\SI{30}{dB}$ and \mbox{$\delta=\SI{0.1}{ppm}$} ({\color[rgb]{0.3922,0.3922,0.3922}\textbf{\textemdash}}), \mbox{$\delta=\SI{1}{ppm}$} ({\color[rgb]{0.1490,0.3569,0.4823}\textbf{\textemdash}}), \mbox{$\delta=\SI{10}{ppm}$} ({\color[rgb]{0.4078,0.3647,0.4745}\textbf{\textemdash}}), \mbox{$\delta=\SI{100}{ppm}$} ({\color[rgb]{0.6706,0.4235,0.5098}\textbf{\textemdash}}), and \mbox{$\delta=\SI{1000}{ppm}$} ({\color[rgb]{0.8471,0.4510,0.4980}\textbf{\textemdash}}). For comparison, the approximate CRLB from and \eqref{eq:SFO_LS_4} ({\color[rgb]{0.8471,0.4510,0.4980}\textbf{\textendash~\textendash}}) and the lower bound for the MLE estimator with ZP from \eqref{eq:sqrt_mle_tau_2} ({\color[rgb]{0.1490,0.3569,0.4824}\textbf{\textendash~\textendash}}) are also shown.}\label{fig:accuracy_delay_sfo}
		\vspace{-0.30cm}
	\end{figure}
	\begin{figure}[!t]
		\centering
		
		\psfrag{55}{(a)}
		\psfrag{22}{(b)}
		
		\psfrag{0}[c][c]{\scriptsize $0$}
		\psfrag{10}[c][c]{\scriptsize $1022$}
		\psfrag{12}[c][c]{\scriptsize $2046$}
		
		\psfrag{0}[c][c]{\scriptsize $0$}
		\psfrag{-15}[c][c]{\scriptsize -$15$}
		\psfrag{-30}[c][c]{\scriptsize -$30$}
		\psfrag{-45}[c][c]{\scriptsize -$45$}
		\psfrag{-60}[c][c]{\scriptsize -$60$}
		
		\psfrag{0}[c][c]{\scriptsize $0$}
		\psfrag{200}[c][c]{\scriptsize $200$}
		\psfrag{400}[c][c]{\scriptsize $400$}
		\psfrag{600}[c][c]{\scriptsize $600$}
		\psfrag{800}[c][c]{\scriptsize $800$}
		
		\psfrag{xxxx}[c][c]{$m_\mathrm{pil}$}
		\psfrag{Rel. bist. range (m)}[c][c]{\small Rel. bist. range (m)}
		\psfrag{Norm. mag. (dB)}[c][c]{\small Norm. mag. (dB)}
		
		\includegraphics[width=8.75cm]{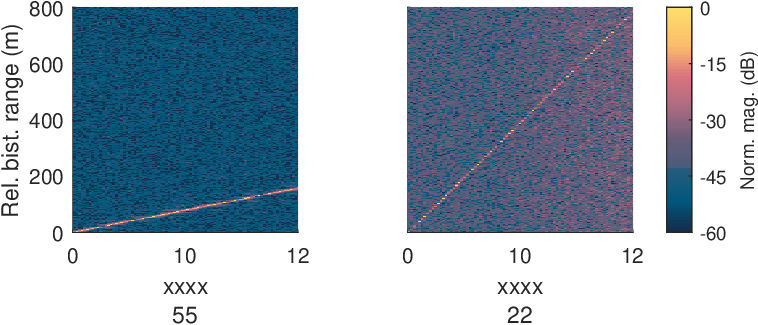}
		\captionsetup{justification=raggedright,labelsep=period,singlelinecheck=false}
		\caption{\ Normalized delay-slow time profile assuming a reference path with $\mathrm{SNR} = \SI{20}{dB}$ and a radar target with $\mathrm{SNR} = \SI{-10}{dB}$ at a relative bistatic range of $\SI{5}{\meter}$ and Doppler shift of $\SI{5}{\kilo\hertz}$ for (a) $\delta=\SI{100}{ppm}$ and (b) $\delta=\SI{500}{ppm}$. While a constant normalized magnitude ratio between the reference path and the noise floor is observed over all $m_\mathrm{pil}$ for $\delta=\SI{100}{ppm}$, a smooth degradation is observed for $\delta=\SI{500}{ppm}$ starting at $m_\mathrm{pil}\approx1022$.}\label{fig:delay_profile}
	\end{figure}

	\subsection{SFO estimation}\label{subsec:sfoEst}

	To perform \ac{SFO} estimation, the approach based on the method by Wu et al.~\cite{wu2012}, which is equivalent to the \ac{TITO} method when forcing $M_\mathrm{pil}^\mathrm{TITO}=M_\mathrm{pil}$, first estimates the delay migration, and then its slope along the pilot \ac{OFDM} symbols. Only then the \ac{SFO} is estimated via \eqref{eq:SFO_LS_2}. Therefore, as discussed in Section~\ref{sec:SFO_sec}, the delay migration estimates must be accurate in order for an accurate \ac{SFO} estimate to be obtained. In this context, the \ac{RMSE} of the \ac{SFO} estimates obtained with the approach based on the method by Wu et al. are shown in Fig.~\ref{fig:accuracy_delay_sfo} for $\delta\in\SI[parse-numbers = false]{\{0.1,1,10,100,1000\}}{ppm}$. These results were obtained from the \ac{SFO} estimates based on delay migration estimation with \ac{ZP} factor $\eta=20$ for $\mathrm{SNR} = \SI{-30}{dB}$ through $\SI{30}{dB}$. %
	To reduce the computational complexity associated with the the high \ac{ZP} factor $\eta$, \ac{CZT} can be used instead of \ac{IDFT} only around the expected reference path tap position in the columns of $\mathbf{D}$  \cite{bhutani2019}.
	
	\begin{figure}[!t]
		\centering
		
		\psfrag{0}[c][c]{\scriptsize $0$}
		\psfrag{511}[c][c]{\scriptsize $511$}
		\psfrag{1023}[c][c]{\scriptsize $1023$}
		\psfrag{1535}[c][c]{\scriptsize $1535$}
		\psfrag{2047}[c][c]{\scriptsize $2047$}
		
		\psfrag{48.3}[c][c]{\scriptsize $48.3$}
		\psfrag{48.5}[c][c]{\scriptsize $48.5$}
		\psfrag{48.7}[c][c]{\scriptsize $48.7$}
		
		\psfrag{0}[c][c]{\scriptsize $0$}
		\psfrag{10}[c][c]{\scriptsize $10$}
		\psfrag{20}[c][c]{\scriptsize $20$}
		\psfrag{30}[c][c]{\scriptsize $30$}
		\psfrag{40}[c][c]{\scriptsize $40$}
		\psfrag{50}[c][c]{\scriptsize $50$}
		
		\psfrag{YYYYYYYYYY}{\small $\mathrm{SINR}$ (dB)}
		\psfrag{XXXX}{\small $m_\mathrm{pil}$}
		
		\includegraphics[width=5.5cm]{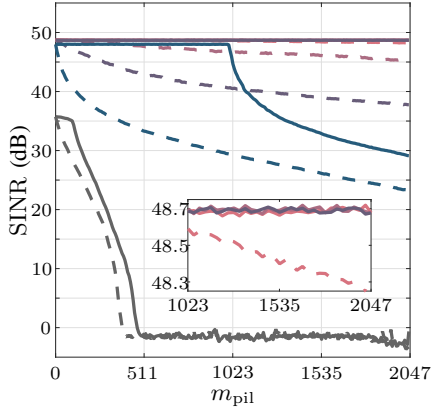}

		\captionsetup{justification=raggedright,labelsep=period,singlelinecheck=false}
		
		\caption{\ SINR of the corresponding peak to the estimated reference path in the \ac{CIR} estimate at the $m_\mathrm{pil}\mathrm{th}$ column of $\mathbf{D}$ used for obtaining the delay migration estimate $\widehat{\Delta\tau}^\mathrm{SFO}_{\mathrm{pil},m_\mathrm{pil}}$. Results are shown for \mbox{$\lvert \delta\rvert=\SI{0.1}{ppm}$ ({\color[rgb]{0.8471,0.4510,0.4980}\textbf{\textemdash}})}, \mbox{$\lvert \delta\rvert=\SI{1}{ppm}$ ({\color[rgb]{0.6706,0.4235,0.5098}\textbf{\textemdash}})}, \mbox{$\lvert \delta\rvert=\SI{10}{ppm}$ ({\color[rgb]{0.4078,0.3647,0.4745}\textbf{\textemdash}})}, \mbox{$\lvert \delta\rvert=\SI{100}{ppm}$ ({\color[rgb]{0.1490,0.3569,0.4824}\textbf{\textemdash}})}, and \mbox{$\lvert \delta\rvert=\SI{1000}{ppm}$ ({\color[rgb]{0.5294,0.5294,0.5294}\textbf{\textemdash}})}, with continuous lines ({\color[rgb]{0,0,0}\textbf{\textemdash}}) for positive $\delta$ and dashed lines for negative $\delta$ ({\color[rgb]{0,0,0}\textbf{\textendash~\textendash}}).}\label{fig:SINR_SFO}
	\end{figure}
	\begin{figure}[!t]
		\centering
		\psfrag{35}[c][c]{\scriptsize $35$}
		\psfrag{37.5}[c][c]{\scriptsize $37.5$}
		\psfrag{40}[c][c]{\scriptsize $40$}
		\psfrag{42.5}[c][c]{\scriptsize $42.5$}
		\psfrag{45}[c][c]{\scriptsize $45$}
		\psfrag{47.5}[c][c]{\scriptsize $47.5$}
		\psfrag{50}[c][c]{\scriptsize $50$}
		
		\psfrag{0}[c][c]{\scriptsize $0$}
		\psfrag{200}[c][c]{\scriptsize $200$}
		\psfrag{400}[c][c]{\scriptsize $400$}
		\psfrag{600}[c][c]{\scriptsize $600$}
		\psfrag{800}[c][c]{\scriptsize $800$}
		\psfrag{1000}[c][c]{\scriptsize $1000$}
		
		\psfrag{YYYYYYYYYY}{\small $\mathrm{SINR}$ (dB)}
		\psfrag{XXXXXXXXXX}{\small $\delta$ (ppm)}
		
		\includegraphics[width=5.5cm]{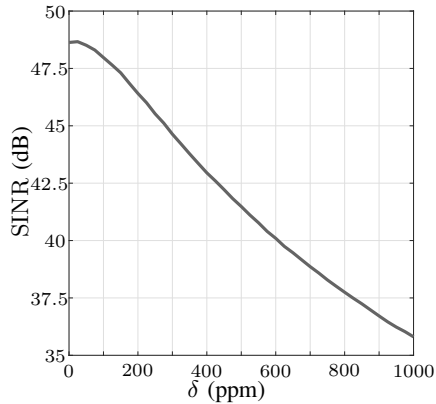}
		
		\captionsetup{justification=raggedright,labelsep=period,singlelinecheck=false}
		\caption{\ SINR as a function of the SFO $\delta$ for the corresponding peak to the estimated reference path in the \ac{CIR} estimate at the first column of $\mathbf{D}$ ($m_\mathrm{pil}=0$), based on which $\widehat{\Delta\tau}^\mathrm{SFO}_{\mathrm{pil},0}$ is estimated.}\label{fig:SINR_1st_symbol_SFO}
	\end{figure}
	
	The obtained \ac{RMSE} results show a tendency of the \ac{SFO} estimates with increasing $\lvert\delta\vert$. The only exception is \mbox{$\delta=\SI{0.1}{ppm}$}, which yields worse \ac{RMSE} than, e.g., \mbox{$\delta=\SI{1}{ppm}$} and \mbox{$\delta=\SI{10}{ppm}$} due to the limited delay estimation accuracy, which becomes more critical due to the small delay migration induced by such a low \ac{SFO}. For all other \ac{SFO} values, the \ac{RMSE} degradation is primarily caused by the \ac{SINR} reduction of the peaks corresponding to the reference path in the delay-slow time profile matrix $\mathbf{D}$ that are used to obtain the required delay migration estimates for \ac{SFO} estimation. This reduction happens due to \ac{SFO}-induced \ac{ICI} and \ac{ISI} as discussed in Section~\ref{subsec:sfoImpConst} and illustrated in Fig.~\ref{fig:delay_profile} assuming the same \ac{SNR} conditions for the reference path and the radar target as for the results in Fig.~\ref{fig:I_full}.

	To analyze the \ac{SINR} degradation more closely, the simulated \ac{SINR} of the reference path estimated with each $m_\mathrm{pil}\mathrm{th}$ pilot \ac{OFDM} symbol, i.e., along the \ac{CIR} estimates in the columns of the delay-slow time profile matrix $\mathbf{D}$, is shown in Fig.~\ref{fig:SINR_SFO} for \mbox{$\mathrm{SNR}=\SI{20}{dB}$} and \mbox{$\delta\in\SI[parse-numbers = false]{\{\pm0.1,\pm1,\pm10,\pm100,\pm1000\}}{ppm}$}. To avoid biasing the \ac{SINR} results, a Chebyshev window with $\SI{100}{dB}$ sidelobe supression was used before forming the delay profiles at each column of $\mathbf{D}$, and the \ac{SINR} was calculated as the ratio between the peak corresponding to the reference path and the region outside its main lobe. The obtained results show that the \ac{SINR} at $m_\mathrm{pil}=0$ decreases with increasing $\lvert\delta\vert$. This is solely due to the \ac{SFO}-induced \ac{ICI} for positive $\delta$, since the \ac{CP} prevents \ac{ISI} due to delay migration at this point, and due to both \ac{SFO}-induced \ac{ICI} and \ac{ISI} for negative $\delta$. A relevant \ac{SINR} decrease, however, is only observed for \mbox{$\lvert\delta\vert=\SI{1000}{ppm}$} among the considered $\delta$ values, while all the other \acp{SFO} result in negligible interference and therefore close \ac{SINR} value to the expected \ac{SNR} of $\SI{48.70}{dB}$. This value is equal to the input \ac{SNR} of $\SI{20}{dB}$, plus the processing gain of  $10\log_{10}(N_\mathrm{pil})=\SI{30.10}{dB}$, minus the windowing \ac{SNR} loss of around $\SI{1.40}{dB}$. To provide a more thorough analysis of the effect of the \ac{SFO}-induced \ac{ICI}, Fig.~\ref{fig:SINR_1st_symbol_SFO} shows the \ac{SINR} at $m_\mathrm{pil}=0$ as a function of $\delta$. Since, as mentioned before, \ac{ISI} already happens at $m_\mathrm{pil}=0$ for negative $\delta$ among the considered values, only positive $\delta$ is analyzed for the presented results in this figure. An \ac{SINR} degradation due to \ac{SFO}-induced \ac{ICI} of $\SI{1}{dB}$ is only experienced at $\SI{125}{ppm}$, which is already slightly above the maximum of $\SI{123}{ppm}$ expected in the setup later considered in Section~\ref{sec:measResults}. Moreover, a \ac{SINR} degradation of $\SI{3}{dB}$ or higher is only experienced for \mbox{$\delta\geq\SI{250}{ppm}$}. This allows concluding that only rather high \acp{SFO} lead to significant \ac{SINR} loss due to the induced \ac{ICI}. 
	Back to Fig.~\ref{fig:SINR_SFO}, it is observed that the \ac{SINR} continuously drop for negative $\delta$ along with $m_\mathrm{pil}$, since all \ac{OFDM} symbols in the frame are affected by the aforementioned \ac{SFO}-induced \ac{ISI}, which becomes more severe with increasing $m_\mathrm{pil}$ as the absolute value of the delay caused by \ac{SFO} from \eqref{eq:SFO_tau_m} increases. Under influence of positive $\delta$, however, the \ac{SINR} only starts to drop after the \ac{SFO}-induced delay exceeds the \ac{CP} duration, only then causing \ac{ISI}.
	
	\begin{figure}[!t]
		\centering
		\psfrag{-1000}[c][c]{\scriptsize -$1000$}
		\psfrag{-500}[c][c]{\scriptsize -$500$}
		\psfrag{0}[c][c]{\scriptsize -$0$}
		\psfrag{500}[c][c]{\scriptsize $500$}
		\psfrag{1000}[c][c]{\scriptsize $1000$}
		
		\psfrag{AAAA}[c][c]{\scriptsize $10^{-8}$}
		\psfrag{BBBB}[c][c]{\scriptsize $10^{-6}$}
		\psfrag{CCCC}[c][c]{\scriptsize $10^{-4}$}
		\psfrag{DDDD}[c][c]{\scriptsize $10^{-2}$}
		\psfrag{EEEE}[c][c]{\scriptsize $10^{0}$}
		\psfrag{FFFF}[c][c]{\scriptsize $10^{2}$}
		\psfrag{GGGG}[c][c]{\scriptsize $10^{4}$}
		
		\psfrag{XXXXXXX}{\small SFO (ppm)}
		\psfrag{YYYYYYYYYYYY}{\small RMSE (ppm)}
		
		\includegraphics[width=5.5cm]{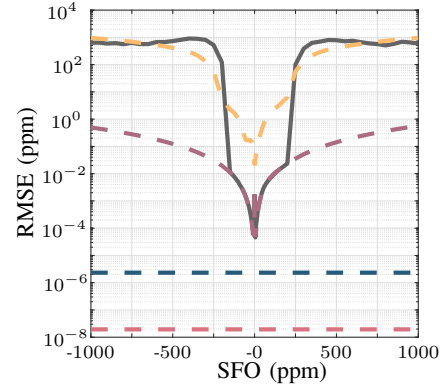}
		
		\captionsetup{justification=raggedright,labelsep=period,singlelinecheck=false}
		
		\caption{\ RMSE of SFO estimates at $\mathrm{SNR} = \SI{20}{dB}$ for $\delta=\SI{-1000}{ppm}$ through $\SI{1000}{ppm}$ with the proposed TITO method ({\color[rgb]{0.6706,0.4235,0.5098}\textbf{\textendash~\textendash}}), the approach based on the method by Wu et al.~\cite{wu2012} ({\color[rgb]{0.5294,0.5294,0.5294}\textbf{\textemdash}}), and the method by Tsai et al.~\cite{tsai2005} ({\color[rgb]{0.9882,0.7333,0.4275}\textbf{\textendash~\textendash}}). For further comparison, the CRLB from \eqref{eq:SFO_LS_4} ({\color[rgb]{0.8471,0.4510,0.4980}\textbf{\textendash~\textendash}}) and the lower bound for the MLE estimator with ZP from \eqref{eq:sqrt_mle_tau_2} ({\color[rgb]{0.1490,0.3569,0.4824}\textbf{\textendash~\textendash}}) are also shown.}\label{fig:RMSE_sfo_full_vs_adaptive_20dBSNR_m1e3_1e3}
	\end{figure}	
	\begin{figure}[!t]
		\centering
		
		\subfloat[ ]{
			\psfrag{-1000}[c][c]{\scriptsize -$1000$}
			\psfrag{-600}[c][c]{\scriptsize -$600$}
			\psfrag{-200}[c][c]{\scriptsize -$200$}
			\psfrag{200}[c][c]{\scriptsize $200$}
			\psfrag{600}[c][c]{\scriptsize $600$}
			\psfrag{1000}[c][c]{\scriptsize $1000$}
			
			\psfrag{8}[c][c]{\scriptsize $8$}
			\psfrag{9}[c][c]{\scriptsize $9$}
			\psfrag{10}[c][c]{\scriptsize $10$}
			\psfrag{11}[c][c]{\scriptsize $11$}
			
			\psfrag{YYYYYYYYYYY}{\small $\log_2(M_\mathrm{pil}^\mathrm{TITO})$}
			\psfrag{XXXXXXXXXX}{\small $\delta$ (ppm)}
			
			\includegraphics[width=5.5cm]{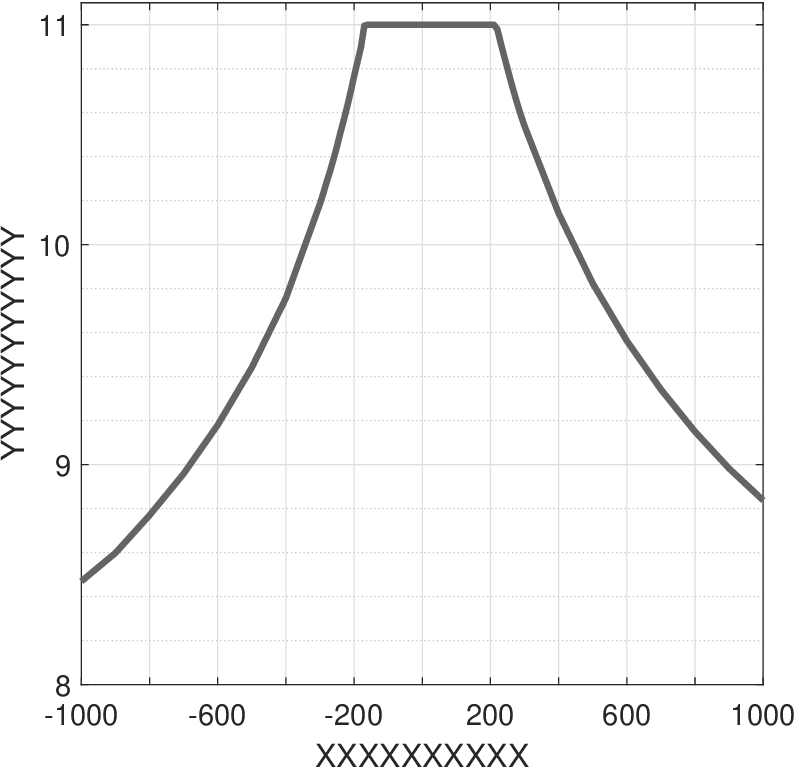}
		}\\
		\subfloat[ ]{
			
			\psfrag{-1000}[c][c]{\scriptsize -$1000$}
			\psfrag{-600}[c][c]{\scriptsize -$600$}
			\psfrag{-200}[c][c]{\scriptsize -$200$}
			\psfrag{200}[c][c]{\scriptsize $200$}
			\psfrag{600}[c][c]{\scriptsize $600$}
			\psfrag{1000}[c][c]{\scriptsize $1000$}
			
			\psfrag{0}[c][c]{\scriptsize $0$}
			\psfrag{20}[c][c]{\scriptsize $20$}
			\psfrag{40}[c][c]{\scriptsize $40$}
			\psfrag{60}[c][c]{\scriptsize $60$}
			\psfrag{80}[c][c]{\scriptsize $80$}
			\psfrag{100}[c][c]{\scriptsize $100$}
			
			\psfrag{YYYYYYYYYYYYYYYYYYYYYYYYYY}{\small Norm. comp. complexity (\%)}
			\psfrag{XXXXXXXXXX}{\small $\delta$ (ppm)}
			
			\includegraphics[width=5.5cm]{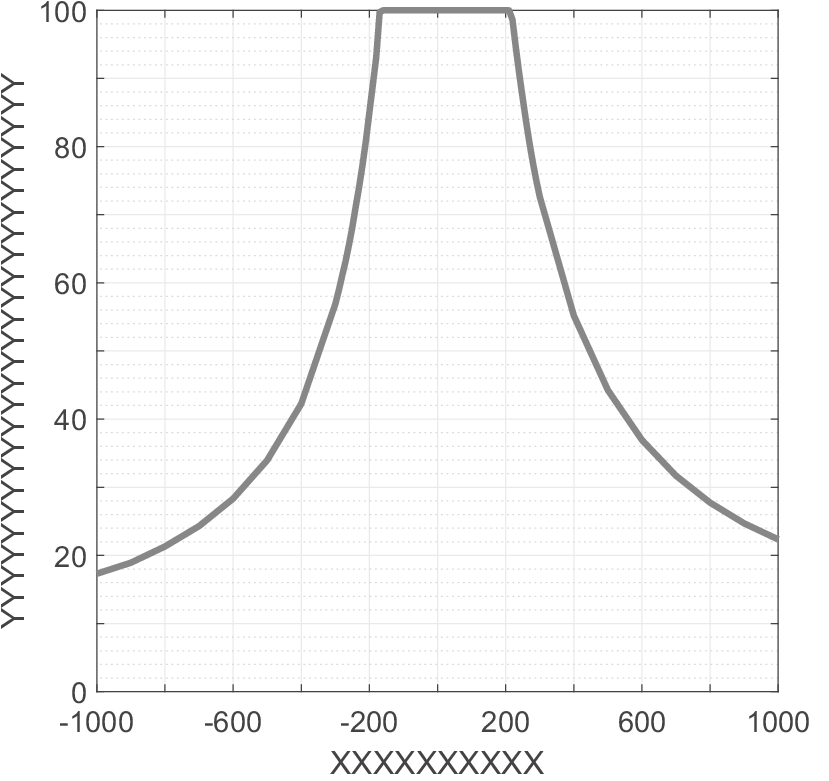}
			
		}

		\captionsetup{justification=raggedright,labelsep=period,singlelinecheck=false}
		
		\caption{\ Reduction in the number of used pilot OFDM symbols for SFO estimation with the TITO method at $\mathrm{SNR}=\SI{20}{dB}$: (a) total number $M_\mathrm{pil}^\mathrm{TITO}$ of used pilot OFDM symbols as a function of the SFO $\delta$ and (b) normalized computational complexity compared with the approach based on the method by Wu et al.~\cite{wu2012}. Between around $\delta=\SI{-150}{ppm}$ and $\delta=\SI{200}{ppm}$ there is a plateau with $M_\mathrm{pil}^\mathrm{TITO}=M_\mathrm{pil}$. For these SFO values, all pilot OFDM symbols are used for the SFO estimation with the TITO method and, consequently, no computational complexity reduction is experienced.}\label{fig:Mpil_max_SFO}
	\end{figure}
		
\begin{figure}[!t]
		\centering
		
		\psfrag{1}[c][c]{\scriptsize $0$}
		\psfrag{3}[c][c]{\scriptsize $3$}
		\psfrag{5}[c][c]{\scriptsize $5$}
		\psfrag{7}[c][c]{\scriptsize $7$}
		\psfrag{9}[c][c]{\scriptsize $9$}
		\psfrag{11}[c][c]{\scriptsize $11$}
		
		\psfrag{AAAA}[c][c]{\scriptsize $10^{-8}$}
		\psfrag{BBBB}[c][c]{\scriptsize $10^{-6}$}
		\psfrag{CCCC}[c][c]{\scriptsize $10^{-4}$}
		\psfrag{DDDD}[c][c]{\scriptsize $10^{-2}$}
		\psfrag{EEEE}[c][c]{\scriptsize $10^{0}$}
		\psfrag{FFFF}[c][c]{\scriptsize $10^{2}$}
		
		\psfrag{YYYYYYYYYYYYY}{\small RMSE (ppm)}
		\psfrag{log2 XXXXXX}{\small $\log_2(M_\mathrm{pil}^\mathrm{TITO})$}
		
		\includegraphics[width=5.5cm]{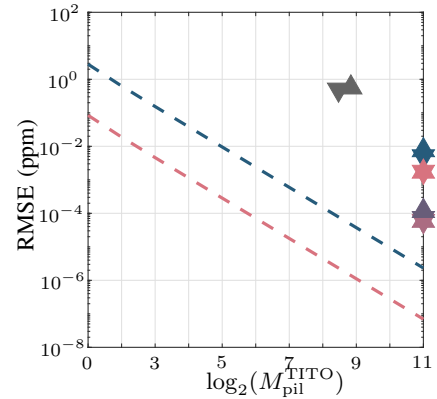}	
		
		\captionsetup{justification=raggedright,labelsep=period,singlelinecheck=false}
		
		\caption{\ RMSE of the SFO estimates at $\mathrm{SNR}=\SI{20}{dB}$ and the total number $M_\mathrm{pil}^\mathrm{TITO}$ of used pilot OFDM symbols for SFO estimation with the TITO method for $\delta=\SI{-1000}{ppm}$ ({\color[rgb]{0.3922,0.3922,0.3922}\textbf{$\blacktriangledown$}}), $\delta=\SI{-100}{ppm}$ ({\color[rgb]{0.1490,0.3569,0.4823}\textbf{$\blacktriangledown$}}), $\delta=\SI{-10}{ppm}$ ({\color[rgb]{0.4078,0.3647,0.4745}\textbf{$\blacktriangledown$}}), $\delta=\SI{-1}{ppm}$ ({\color[rgb]{0.6706,0.4235,0.5098}\textbf{$\blacktriangledown$}}), $\delta=\SI{-0.1}{ppm}$ ({\color[rgb]{0.8471,0.4510,0.4980}\textbf{$\blacktriangledown$}}), $\delta=\SI{0.1}{ppm}$ ({\color[rgb]{0.8471,0.4510,0.4980}\textbf{$\blacktriangle$}}), $\delta=\SI{1}{ppm}$ ({\color[rgb]{0.6706,0.4235,0.5098}\textbf{$\blacktriangle$}}), $\delta=\SI{10}{ppm}$ ({\color[rgb]{0.4078,0.3647,0.4745}\textbf{$\blacktriangle$}}), $\delta=\SI{100}{ppm}$ ({\color[rgb]{0.1490,0.3569,0.4823}\textbf{$\blacktriangle$}}), and $\delta=\SI{1000}{ppm}$ ({\color[rgb]{0.3922,0.3922,0.3922}\textbf{$\blacktriangle$}}). For comparison, the CRLB from \eqref{eq:SFO_LS_4} ({\color[rgb]{0.8471,0.4510,0.4980}\textbf{\textendash~\textendash}}) and the lower bound for the MLE estimator with ZP from \eqref{eq:sqrt_mle_tau_2} ({\color[rgb]{0.1490,0.3569,0.4824}\textbf{\textendash~\textendash}}) as functions of $M_\mathrm{pil}^\mathrm{TITO}$ are also shown.}\label{fig:CRLB_SFO_Mpil}
	\end{figure}

	The discussed results so far have indicated that the \ac{SFO} estimates with the approach based on the method by Wu et al. may be significantly biased even without relevant influence of \ac{SFO}-induced \ac{ICI} when \ac{ISI} eventually happens, as shown in Fig.~\ref{fig:SINR_SFO}. This is caused by the use of the \ac{CIR} estimates in the columns of the delay-slow time profile matrix $\mathbf{D}$ that correspond to pilot \ac{OFDM} symbols of index $m_\mathrm{pil}$ that are affected by \ac{ISI}. As discussed in Section~\ref{subsec:SFO_TITO_est}, this is circumvented with the \ac{TITO} method by setting an upperbound to the step of the reference path delay migration between two consecutive estimates. Once this threshold is exceeded, the corresponding pilot \ac{OFDM} symbol $m_\mathrm{pil}$ and all next ones are assumed to be under strong \ac{SFO}-induced \ac{ISI} and are therefore disregarded, leading to a total number of used pilot \ac{OFDM} symbols $M_\mathrm{pil}^\mathrm{TITO}\leq M_\mathrm{pil}$ for \ac{SFO} estimation with the \ac{TITO} method. Assuming \mbox{$\mathrm{SNR}=\SI{20}{dB}$} as in the previous analysis in this section, Fig.~\ref{fig:RMSE_sfo_full_vs_adaptive_20dBSNR_m1e3_1e3} shows the achieved \ac{RMSE} of
	the \ac{SFO} estimates with both the approach based on the method by Wu et al. and the \ac{TITO} method assuming \mbox{$\delta_\mathrm{max}=\SI{1000}{ppm}$} and \mbox{$\varepsilon=0.1$} for \mbox{$\delta=\SI{-1000}{ppm}$} through \mbox{$\delta=\SI{1000}{ppm}$}. For comparison, the performance of the method by Tsai et al.~\cite{tsai2005} is also shown. The achieved results show that, between around \mbox{$\delta=\SI{-150}{ppm}$} and \mbox{$\delta=\SI{200}{ppm}$}, \ac{TITO} yields equivalent performance to the delay-based method based on Wu et al., while the method by Tsai et al. performs the worst. In this \ac{SFO} region, the experienced processing gain of \mbox{$10\log_{10}(N_\mathrm{pil})=\SI{30.10}{dB}$} when analyzing the reference path in the obtained the \ac{CIR} estimates stored in the columns of $\mathbf{D}$ in the former two methods is sufficient to avoid severe degradation of the reference path \ac{SINR} and consequently of the \ac{SFO} estimates due to the \ac{SFO}-induced \ac{ISI}. The method by Tsai et al., however, does not profit from this processing gain as it operates on the subcarrier level. Outside this \ac{SFO} region, the \ac{RMSE} of the method based on Wu et al. is significantly worsened by the \ac{ISI}-contamined pilot \ac{OFDM} symbols used to estimate the aforementioned \acp{CIR}. While the Tsai method achieves equal or better performance, it is also affected by \ac{SFO}-induced \ac{ISI} and its \ac{RMSE} is still rather high, i.e., $\SI{5.6}{ppm}$ or more. The \ac{TITO} method, in turn, circumvents the \ac{ISI} issue with its choice of $M_\mathrm{pil}^\mathrm{TITO}\leq M_\mathrm{pil}$ pilot \ac{OFDM} symbols used for the delay migration estimation required to estimate the \ac{SFO} as described by \eqref{eq:M_TITO} and illustrated in Fig.~\ref{fig:Mpil_max_SFO}a. In Fig.~\ref{fig:Mpil_max_SFO}b, the resulting computational complexity experienced as less pilot \ac{OFDM} symbols are processed at high \ac{SFO} values is shown. The obtained results allow concluding that the performance of \ac{TITO} is mainly affected by the \ac{SFO}-induced \ac{ICI} and the reduced number of used pilot \ac{OFDM} symbols, which affects the \ac{SFO} estimation accuracy as illustrated in Fig.~\ref{fig:CRLB_SFO_Mpil}. Still, \ac{TITO} outperforms both its counterparts for the adopted \ac{OFDM} signal parameters, presenting \acp{RMSE} as low as $\SI{0.01}{ppm}$ for $\lvert\delta\rvert=\SI{150}{ppm}$ and around $\SI{0.5}{ppm}$ for $\lvert\delta\rvert=\SI{1000}{ppm}$.
	
	\begin{figure}[!t]
		\centering
		
		\includegraphics[width=8.5cm]{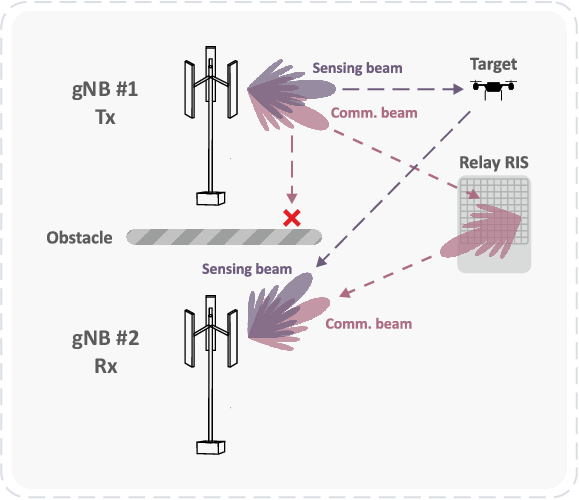}
		\captionsetup{justification=raggedright,labelsep=period,singlelinecheck=false}
		\caption{\ RIS-assisted bistatic ISAC system concept.}\label{fig:systemModel_gNB_RIS}	
		
	\end{figure}
	
	\begin{figure*}[!t]
		\centering
		\subfloat[ ]{
			\includegraphics[width=5.75cm]{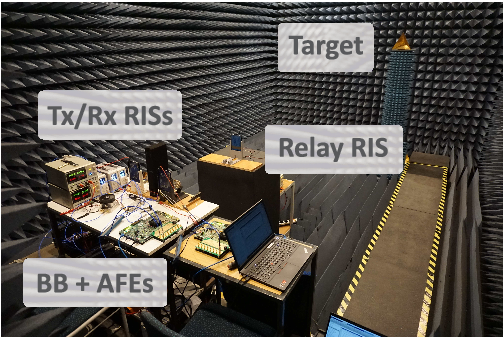}
		}
		\subfloat[ ]{
			\includegraphics[width=5.75cm]{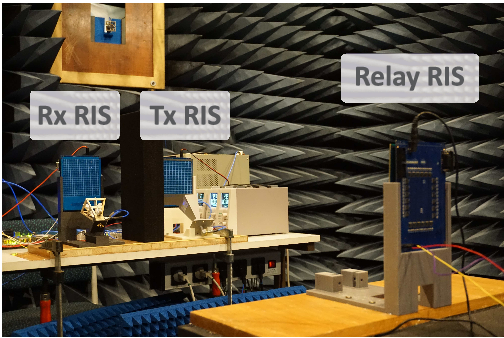}
		}
		\subfloat[ ]{
			\includegraphics[width=5.75cm]{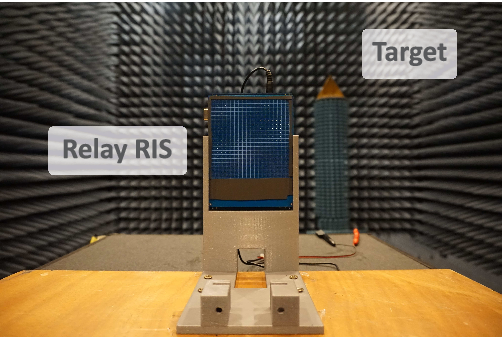}
		}
		\captionsetup{justification=raggedright,labelsep=period,singlelinecheck=false}
		\caption{\ Multiple views of the assembled measurement setup for the RIS-based bistatic ISAC system in the IHE anechoic chamber at the KIT.}\label{fig:measSetup}
	\end{figure*}	
	\begin{figure}[!t]
		\centering
		
		\psfrag{55}{(a)}
		\psfrag{22}{(b)}
		\psfrag{33}{(c)}
		
		\psfrag{-2}[c][c]{\scriptsize -$2$}
		\psfrag{-1}[c][c]{\scriptsize -$1$}
		\psfrag{0}[c][c]{\scriptsize $0$}
		\psfrag{1}[c][c]{\scriptsize $1$}
		\psfrag{2}[c][c]{\scriptsize $2$}
		
		\psfrag{10-4}[c][c]{\scriptsize -$4$}
		\psfrag{10-3}[c][c]{\scriptsize -$3$}
		\psfrag{10-2}[c][c]{\scriptsize -$2$}
		\psfrag{10-1}[c][c]{\scriptsize -$1$}
		\psfrag{10-0}[c][c]{\scriptsize \hspace{-.2cm}$0$}
		
		\psfrag{I}{\small $I$}
		\psfrag{Q}{\small $Q$}
		
		\psfrag{Norm. density}[c][c]{\scriptsize $\log_{10}(\mathrm{Norm. density})$}
		
		\includegraphics[width=\columnwidth]{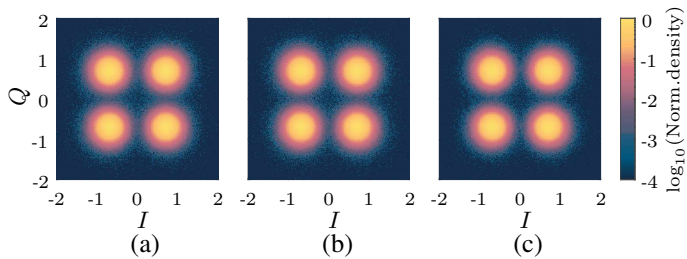}
		\captionsetup{justification=raggedright,labelsep=period,singlelinecheck=false}
		\caption{\ Measured QPSK constellations assuming SFO estimation and correction based on (a) the method by Tsai et al. without residual SFO correction, (b) the method by Tsai et al. with residual SFO correction, and (c) the TITO method.}\label{fig:meas_constDensity}
	\end{figure}
		
	\section{Measurement Setup and Results}\label{sec:measResults}
	
	In this section, the benefits of the proposed \ac{TITO} method for \ac{SFO} estimation and subsequent correction are demonstrated in practice.  This is done via measurements based on the \ac{RIS}-assisted bistatic \ac{ISAC} system concept depicted in Fig.~\ref{fig:systemModel_gNB_RIS}, which is proposed in this article to overcome \ac{NLoS} conditions. In this concept, it is assumed that the \ac{LoS} path between the \acp{gNB} is obstructed, and that a relay \ac{RIS} is used to steer the incoming signals from \ac{gNB} \#1 in the directions of \ac{gNB} \#2 \cite{li2021}.
	
	The corresponding measurement setup to the proposed \ac{RIS}-assisted  bistatic \ac{ISAC} system concept shown in Fig.~\ref{fig:measSetup} was assembled in the anechoic chamber of the \ac{IHE} at the \ac{KIT}. Its \ac{BB} part consisted of two Zynq UltraScale+ RFSoC ZCU111 \ac{SoC} platforms from Xilinx, Inc, one acting as a transmitter and another one as a receiver. Both the transmitter and receiver boards were connected to the same \ac{RF} \acp{AFE} used in \cite{li2021} via coaxial cables, which then transmit/receive \ac{OFDM} signals with the parameters listed in Table~\ref{tab:ofdmParameters} to/from the transmit and receive \acp{RIS} that were used as the arrays of \ac{gNB} \#1 and \#2, respectively. Since the \ac{LoS} link between transmit and receive \acp{RIS} was obstructed by an \ac{RF} absorber, a third \ac{RIS} with known position was used as a relay. Beamforming was then performed for transmit and receive \acp{RIS} so that they both had one beam pointing at the relay \acp{RIS} and another beam pointing at a corner reflector placed at the other side of the anechoic chamber to serve as a radar target. While it is reasonable to assume that the beamforming directions towards the relay \ac{RIS} are known, the exact beamforming directions from transmit and receive \acp{RIS} towards the radar target were defined via exhaustive search. In practical deployments, more efficient approaches such as the ones from \cite{li2024,mandelli2023} can be adapted to the bistatic sensing case to both perform initial target detection as mentioned in Section~\ref{sec:system_model}.
			
	A coordinate system was adopted for the measurement setup shown in Fig.~\ref{fig:measSetup} assuming that the origin \mbox{$\SI[parse-numbers = false]{[0,0,0]}{\meter}$} was between the phase center of the transmitter and receiver \acp{RIS}, which were located at \mbox{$\SI[parse-numbers = false]{[-0.15,0,0]}{\meter}$} and \mbox{$\SI[parse-numbers = false]{[0.15,0,0]}{\meter}$}, respectively. Furthermore, the phase center of the relay \ac{RIS} was located at \mbox{$\SI[parse-numbers = false]{[0,1,0]}{\meter}$} and the center of the corner reflector at \mbox{$\SI[parse-numbers = false]{[0.75, 4.50, 0.9]}{\meter}$}. Considering that the reference atennas of both transmitter and receiver were located at a distance of $\SI{0.15}{\meter}$ of their respective \acp{RIS}, the bistatic range ground truth values for the reference path enabled by the relay \ac{RIS} and the radar target were $\SI{2.32}{\meter}$ and $\SI{9.63}{\meter}$, respectively. Assuming perfect synchronization w.r.t. the reference path, this should result in a relative bistatic range of $\SI{7.31}{\meter}$ for the radar target.
	
	\begin{figure}[!t]
		\centering
		
		\psfrag{55}{(a)}
		\psfrag{22}{(b)}
		\psfrag{33}{(c)}
		
		\psfrag{0}[c][c]{\scriptsize $0$}
		\psfrag{5}[c][c]{\scriptsize $5$}
		\psfrag{10}[c][c]{\scriptsize $10$}
		\psfrag{15}[c][c]{\scriptsize $15$}	
		
		\psfrag{-2}[c][c]{\scriptsize -$2$}
		\psfrag{-1}[c][c]{\scriptsize -$1$}
		\psfrag{0}[c][c]{\scriptsize $0$}
		\psfrag{1}[c][c]{\scriptsize $1$}
		\psfrag{2}[c][c]{\scriptsize $2$}
		
		\psfrag{0}[c][c]{\scriptsize $0$}
		\psfrag{-20}[c][c]{\scriptsize -$15$}
		\psfrag{-40}[c][c]{\scriptsize -$30$}
		\psfrag{-60}[c][c]{\scriptsize -$45$}
		\psfrag{-80}[c][c]{\scriptsize -$60$}
		
		\psfrag{Doppler shift (kHz)}[c][c]{\scriptsize Doppler shift (kHz)}
		\psfrag{Rel. bist. range (m)}[c][c]{\scriptsize Rel. bist. range (m)}
		\psfrag{Norm. mag. (dB)}[c][c]{\scriptsize Norm. mag. (dB)}
		
		\includegraphics[width=\columnwidth]{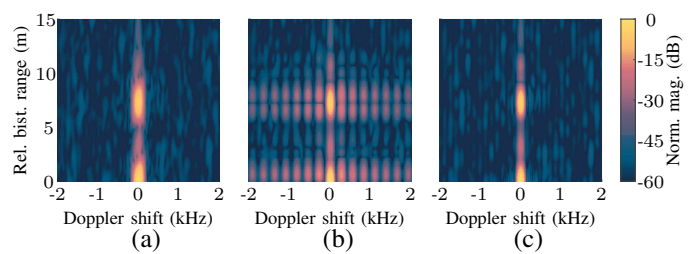}
		\captionsetup{justification=raggedright,labelsep=period,singlelinecheck=false}
		\caption{\ Measured bistatic radar images assuming SFO estimation and correction based on (a) the method by Tsai et al. without residual SFO correction, (b) the method by Tsai et al. with residual SFO correction, and (c) the TITO method.}\label{fig:I_meas}
	\end{figure}
	
	Using the described measurement setup under the aforementioned conditions, 2 preamble \ac{OFDM} symbols were transmitted before the \ac{OFDM} frame to enable coarse time and frequency synchronization with the \ac{SC} algorithm \cite{schmidl1997}. Afterwards, a fine time synchronization was performed via cross-correlation with the first preamble \ac{OFDM} symbol to obtain a more accurate estimate of the \ac{OFDM} frame start sample. After this point, \ac{SFO} estimation was performed. Using the \ac{TITO} method with \mbox{$\delta_\mathrm{max}=\SI{123}{ppm}$}, which is equal to the absolute value of maximum expected \ac{SFO} in the adopted setup, the same \ac{SFO} estimate of \mbox{$\SI{-104.29}{ppm}$} was obtained as with the algorithm based on Wu et al.~\cite{wu2012}. This was the case since only tolerable \ac{SINR} degradation due to \ac{SFO}-induced \ac{ICI} and \ac{ISI} was experienced, and therefore \eqref{eq:M_TITO} resulted in \mbox{$M_\mathrm{pil}^\mathrm{TITO}=M_\mathrm{pil}$} pilot \ac{OFDM} symbols used by the \ac{TITO} method to estimate the \ac{SFO} with \eqref{eq:SFO_LS_2}, meaning that both methods are equivalent for the adopted set of parameters and hardware. Since, however, the \ac{TITO} method achieves equal or better performance compared to the method based on Wu et al.~\cite{wu2012}, the latter will be excluded from the subsequent analyses. The method by Tsai et al.~\cite{tsai2005}, however, yielded a biased \ac{SFO} estimate of $\SI{11.60}{ppm}$, which ultimatelly resulted in a \ac{BER} of $0.41$ even with \ac{LDPC} decoding and produced a bistatic radar image where no targets could be detected. Out of this reason, the method by Tsai was alternatively implemented with $20$ dedicated preamble \ac{OFDM} symbols for \ac{SFO} estimation transmitted right before the \ac{OFDM} frame. Although this incurs in intolerable overhead in communication operations, only so an \ac{SFO} estimate of \mbox{$\SI{-103.73}{ppm}$} could be obtained. Since the \ac{SFO} estimation with the Tsai algorithm based on preamble symbols is not accurate enough for bistatic sensing, the latter method was also combined with a residual \ac{SFO} estimation and subsequent correction via discrete-time domain shifting of the \ac{OFDM} symbols as described in \cite{giroto2023_EuMW,burmeister2021}. While the aforementioned residual \ac{SFO} correction compensates the delay migration, it does not alleviate the effect of the \ac{SFO}-induced \ac{ICI}.
	
	After \ac{SFO} estimation and correction via resampling, besides further communication processing as described in \cite{giroto2023_EuMW,brunner2024}, the \ac{QPSK} constellations shown in Fig.~\ref{fig:meas_constDensity} were obtained. The obtained results show virtually the same performance regardless of the \ac{SFO} estimation method. This is evidenced by the \ac{EVM} with mean value of $\SI{-14.84}{dB}$ and standard deviation of $\SI{5.76}{dB}$ obtained in the case where \ac{SFO} estimation was performed with the \ac{TITO} method, while mean \ac{EVM} of $\SI{-14.69}{dB}$ and standard deviation of $\SI{5.79}{dB}$ was obtained for the case where \ac{SFO} estimation was done with the method by Tsai et al. without residual \ac{SFO} correction and mean \ac{EVM} of $\SI{-14.62}{dB}$ and standard deviation of $\SI{5.79}{dB}$ for \ac{SFO} estimation done with the method by Tsai et al. with residual \ac{SFO} correction. In all cases, no transmission errors were experienced due to the chosen \ac{LDPC} channel code \cite{miao2024} of rate $2/3$ and the parity check matrix for the case with $64800$ bits in an \ac{LDPC} code block from \cite{ETSI302307}.
		
	While the \ac{EVM} performance remains virtually unaffected by the choice between the three considered \ac{SFO} estimation and correction methods, the bistatic sensing performance is significantly influenced by residual \ac{SFO} that lead to range and Doppler shift migrations that can be derived from delay and frequency shift migrations in \eqref{eq:SFO_delta_tau_m} and \eqref{eq:SFO_freq_mig}, respectively. Fig.~\ref{fig:I_meas} shows the measured bistatic radar images ultimatelly obtained when using the aforementioned \ac{SFO} estimation and correction approaches. As in \cite{giroto2023_EuMW,brunner2024}, the  required full transmit frame for the bistatic radar processing was reconstructed based on the knowledge of the pilot subcarriers as well as the receive payload \ac{QPSK} symbols from the constellations in Fig.~\ref{fig:meas_constDensity}. Furthermore, biases to the relative bistatic ranges and Doppler shifts in the radar image are removed by shifting the radar image so that the reference path, which is the stronger peak in the radar image, appears at a relative bistatic range of $\SI{0}{\meter}$ and Doppler shift of $\SI{0}{\hertz}$. It is observed that the bistatic radar image obtained after \ac{SFO} estimation with the \ac{TITO} method presents a well-defined radar target peak at a relative bistatic range of $\SI{7.19}{\meter}$. In the radar image obtained after synchronization with the method by Tsai et al. without residual \ac{SFO} correction, the radar target peak appears at the same range as in the previous case and a negligible \ac{SNR} loss of $\SI{0.38}{dB}$ is observed. This only happens due to the compensation of range migration via discrete-time domain shifting of the \ac{OFDM} symbols in the frame. However, this comes at the cost of extra computational complexity and a significant spread in the Doppler shift direction, with multiple high sidelobes that could not be successfully suppressed by windowing. As for the case where residual \ac{SFO} correction was performed after the estimation with the Tsai algorithm and resampling,  a significant degradation of the range resolution and a slight degradation of the Doppler shift resolution are observed compared with the bistatic radar image obtained when the \ac{TITO} method is used. In this case, the radar target peak is at a relative bistatic range of $\SI{7.34}{\meter}$ and a \ac{SNR} loss of $\SI{2.65}{dB}$ is experienced w.r.t. the \ac{TITO} case.


	\section{Conclusion}\label{sec:conclusion}
	
	This article introduced a novel method to perform \ac{ISI}-robust \ac{SFO} estimation towards robust bistatic sensing in \ac{OFDM}-based \ac{ISAC} systems. The proposed method, \ac{TITO}, consists of using pilot \ac{OFDM} subcarriers to obtain \ac{CIR} estimates, from which the delay migration of a dominant, reference path with constant range is estimated. By estimating the slope of the delay migration of the reference path along consecutive \ac{CIR} estimates, the \ac{SFO} itself can then be estimated.
	
	After a mathematical derivation of \ac{SFO} estimation accuracy lower bounds for the cases where the delay migration estimates are performed with unbiased or \ac{MLE} algorithms, the performance of the proposed \ac{TITO} method for \ac{SFO} estimation was verified and compared to alternative techniques in the literature. The achieved results allow concluding that \ac{TITO} offers sufficiently accurate \ac{SFO} estimates even for much higher \ac{SFO} values that are commonly expected in practical deployments. This allows ultimately producing bistatic radar images where target reflections have higher \ac{SNR} and lower sidelobe level than with the methods by Wu et al.~\cite{wu2012} and Tsai et al.~\cite{tsai2005}, enabling
	enabling virtually unbiased radar target parameter estimation. In addition, the computational complexity associated with the \ac{TITO} algorithm is considerably reduced when compared to the its best performing competitor, i.e., the approach based on the method by Wu et al.~\cite{wu2012}, since a reduced number of pilot \ac{OFDM} symbols is processed.
	
	Considering \ac{NLoS} condition between the transmitting and receiving nodes of the bistatic \ac{ISAC} pair, a system concept where the aforementioned path is created by using a relay \ac{RIS} was proposed. This allowed to demonstrate the performance of the \ac{TITO} method in practice, showing that it achieves sufficient \ac{SFO} estimation accuracy to enable communication and radar sensing in the considered bistatic \ac{ISAC} system.
	
	Although not addressed in this article, the proposed \ac{TITO} method for \ac{SFO} estimation can be applied with minor adjustments to bistatic \ac{ISAC} systems based on alternative waveforms such as orthogonal chirp-division multiplexing and phase-modulated continuous wave \cite{giroto2021_tmtt}, being combined to existing time and frequency synchronization techniques \cite{filomeno2022,giroto2023_EuMW_PMCW} to enable robust communication and unbiased bistatic sensing.

	\bibliographystyle{IEEEtran}
	\bibliography{main}

\begin{thebibliography}{10}
\providecommand{\url}[1]{#1}
\csname url@samestyle\endcsname
\providecommand{\newblock}{\relax}
\providecommand{\bibinfo}[2]{#2}
\providecommand{\BIBentrySTDinterwordspacing}{\spaceskip=0pt\relax}
\providecommand{\BIBentryALTinterwordstretchfactor}{4}
\providecommand{\BIBentryALTinterwordspacing}{\spaceskip=\fontdimen2\font plus
\BIBentryALTinterwordstretchfactor\fontdimen3\font minus
  \fontdimen4\font\relax}
\providecommand{\BIBforeignlanguage}[2]{{%
\expandafter\ifx\csname l@#1\endcsname\relax
\typeout{** WARNING: IEEEtran.bst: No hyphenation pattern has been}%
\typeout{** loaded for the language `#1'. Using the pattern for}%
\typeout{** the default language instead.}%
\else
\language=\csname l@#1\endcsname
\fi
#2}}
\providecommand{\BIBdecl}{\relax}
\BIBdecl

\bibitem{viswanathan2020}
H.~Viswanathan and P.~E. Mogensen, ``Communications in the {6G} era,''
  \emph{IEEE Access}, vol.~8, pp. 57\,063--57\,074, Mar. 2020.

\bibitem{lima2021}
C.~{de Lima et al.}, ``Convergent communication, sensing and localization in
  {6G} systems: An overview of technologies, opportunities and challenges,''
  \emph{IEEE Access}, vol.~9, pp. 26\,902--26\,925, Jan. 2021.

\bibitem{wild2021}
T.~{Wild}, V.~{Braun}, and H.~{Viswanathan}, ``Joint design of communication
  and sensing for beyond {5G} and {6G} systems,'' \emph{IEEE Access}, vol.~9,
  pp. 30\,845--30\,857, Feb. 2021.

\bibitem{liu2022}
F.~{Liu et al.}, ``Integrated sensing and communications: Towards
  dual-functional wireless networks for {6G} and beyond,'' \emph{IEEE J. Sel.
  Areas Commun.}, vol.~40, no.~6, pp. 1728--1767, Jun. 2022.

\bibitem{mandelli2023survey}
S.~Mandelli, M.~Henninger, M.~Bauhofer, and T.~Wild, ``Survey on integrated
  sensing and communication performance modeling and use cases feasibility,''
  in \emph{2023 2nd Int. Conf. 6G Netw.}, Oct. 2023, pp. 1--8.

\bibitem{zhang2021}
A.~Zhang, M.~L. Rahman, X.~Huang, Y.~J. Guo, S.~Chen, and R.~W. Heath,
  ``Perceptive mobile networks: Cellular networks with radio vision via joint
  communication and radar sensing,'' \emph{IEEE Veh. Technol. Mag.}, vol.~16,
  no.~2, pp. 20--30, 2021.

\bibitem{kadelka2023}
A.~Kadelka, G.~Zimmermann, J.~Plach\'y, and O.~Holschke, ``A {CSP}'s view on
  opportunities and challenges of integrated communications and sensing,'' in
  \emph{2023 IEEE 3rd Int. Symp. Joint Commun. Sens.}, Mar. 2023, pp. 1--6.

\bibitem{shatov2024}
V.~{Shatov et al.}, ``Joint radar and communications: Architectures, use cases,
  aspects of radio access, signal processing, and hardware,'' \emph{IEEE
  Access}, pp. 1--1, 2024.

\bibitem{barneto2019}
C.~B. {Barneto et al.}, ``Full-duplex {OFDM} radar with {LTE} and {5G} {NR}
  waveforms: Challenges, solutions, and measurements,'' \emph{IEEE Trans.
  Microw. Theory Tech.}, vol.~67, no.~10, pp. 4042--4054, Oct. 2019.

\bibitem{barneto2021}
C.~B. {Barneto}, S.~D. {Liyanaarachchi}, M.~{Heino}, T.~{Riihonen}, and
  M.~{Valkama}, ``Full duplex radio/radar technology: The enabler for advanced
  joint communication and sensing,'' \emph{IEEE Wireless Commun.}, vol.~28,
  no.~1, pp. 82--88, Feb. 2021.

\bibitem{ksiezyk2023}
A.~{Ksi{\k{e}}żyk et al.}, ``Opportunities and limitations in radar sensing
  based on {5G} broadband cellular networks,'' \emph{IEEE Aerosp. Electron.
  Syst. Mag.}, vol.~38, no.~9, pp. 4--21, Sept. 2023.

\bibitem{thomae2023}
R.~Thomä and T.~Dallmann, ``Distributed {ISAC} systems – multisensor radio
  access and coordination,'' in \emph{2023 20th Eur. Radar Conf.}, Sept. 2023,
  pp. 351--354.

\bibitem{bauhofer2023}
M.~Bauhofer, S.~Mandelli, M.~Henninger, T.~Wild, and S.~{ten Brink},
  ``Multi-target localization in multi-static integrated sensing and
  communication deployments,'' in \emph{2023 2nd Int. Conf. 6G Netw.}, Oct.
  2023, pp. 1--4.

\bibitem{samczynski2022}
P.~{Samczy\'{n}ski et al.}, ``{5G} network-based passive radar,'' \emph{IEEE
  Transactions on Geoscience and Remote Sensing}, vol.~60, pp. 1--9, Dec. 2022.

\bibitem{abratkiewicz2023}
K.~Abratkiewicz, A.~Ksi{\k{e}}żyk, M.~P\l{}otka, P.~Samczy\'{n}ski,
  J.~Wszo\l{}ek, and T.~P. Zieli\'{n}ski, ``{SSB}-based signal processing for
  passive radar using a {5G} network,'' \emph{IEEE Journal Sel. Topics Appl.
  Earth Observations Remote Sens.}, vol.~16, pp. 3469--3484, 2023.

\bibitem{giroto2023_EuMW}
L.~{Giroto de Oliveira et al.}, ``Bistatic {OFDM}-based joint
  radar-communication: Synchronization, data communication and sensing,'' in
  \emph{2023 20th Eur. Radar Conf.}, Sept. 2023, pp. 359--362.

\bibitem{giroto2021_tmtt}
L.~{Giroto de Oliveira}, B.~{Nuss}, M.~B. {Alabd}, A.~{Diewald}, M.~{Pauli},
  and T.~{Zwick}, ``Joint radar-communication systems: Modulation schemes and
  system design,'' \emph{IEEE Trans. Microw. Theory Tech.}, vol.~70, no.~3, pp.
  1521--1551, Mar. 2022.

\bibitem{werbunat2024}
D.~{Werbunat et al.}, ``On the synchronization of uncoupled multistatic {PMCW}
  radars,'' \emph{IEEE Trans. Microw. Theory Tech. (Early Access)}, pp. 1--13,
  Feb. 2024.

\bibitem{tsai2005}
P.-Y. {Tsai}, H.-Y. {Kang}, and T.-D. {Chiueh}, ``Joint weighted least-squares
  estimation of carrier-frequency offset and timing offset for {OFDM} systems
  over multipath fading channels,'' \emph{IEEE Trans. Veh. Technol.}, vol.~54,
  no.~1, pp. 211--223, Jan. 2005.

\bibitem{burmeister2021}
F.~Burmeister, R.~Jacob, A.~Traßl, N.~Schwarzenberg, and G.~Fettweis,
  ``Dealing with fractional sampling time offsets for unsynchronized mobile
  channel measurements,'' \emph{IEEE Wireless Commun. Lett}, vol.~10, no.~12,
  pp. 2781--2785, Dec. 2021.

\bibitem{wu2012}
Y.~Wu, Y.~Zhao, and D.~Li, ``Sampling frequency offset estimation for
  pilot-aided {OFDM} systems in mobile environment,'' \emph{Wireless Pers.
  Commun.}, vol.~62, pp. 215--226, Jan. 2012.

\bibitem{dantas2016}
C.~F. Dantas, D.~Castro, and C.~M. Panazio, ``On enhancing the pilot-aided
  sampling clock offset estimation of mobile {OFDM} systems,'' \emph{J. Commun.
  Inf. Sys.}, vol.~31, no.~1, pp. 108--117, Jun. 2016.

\bibitem{pegoraro2024}
J.~{Pegoraro et al.}, ``{JUMP}: Joint communication and sensing with
  unsynchronized transceivers made practical,'' \emph{IEEE Trans. Wireless
  Commun. (Early Access)}, pp. 1--16, Feb. 2024.

\bibitem{farrow1988}
C.~Farrow, ``A continuously variable digital delay element,'' in \emph{1988
  IEEE Int. Symp. Circuits Syst.}, vol.~3, Jun. 1988, pp. 2641--2645.

\bibitem{bjornsson2022}
E.~Bj\"ornson, H.~Wymeersch, B.~Matthiesen, P.~Popovski, L.~Sanguinetti, and
  E.~de~Carvalho, ``Reconfigurable intelligent surfaces: A signal processing
  perspective with wireless applications,'' \emph{IEEE Signal Process. Mag.},
  vol.~39, no.~2, pp. 135--158, Mar. 2022.

\bibitem{chepuri2023}
S.~P. Chepuri, N.~Shlezinger, F.~Liu, G.~C. Alexandropoulos, S.~Buzzi, and
  Y.~C. Eldar, ``Integrated sensing and communications with reconfigurable
  intelligent surfaces: From signal modeling to processing,'' \emph{IEEE Signal
  Processing Magazine}, vol.~40, no.~6, pp. 41--62, Sept. 2023.

\bibitem{elbir2023}
A.~M. Elbir, K.~V. Mishra, M.~R.~B. Shankar, and S.~Chatzinotas, ``The rise of
  intelligent reflecting surfaces in integrated sensing and communications
  paradigms,'' \emph{IEEE Network (Early Access)}, vol.~37, no.~6, pp.
  224--231, Nov. 2023.

\bibitem{skolnik2008}
N.~J. Willis, ``Bistatic radar,'' in \emph{Radar Handbook}, M.~I. Skolnik,
  Ed.\hskip 1em plus 0.5em minus 0.4em\relax New York, NY: The McGraw-Hill
  Companies, 2008, ch.~23, pp. 23.14--23.15.

\bibitem{li2021_search}
Y.~{Li et al.}, ``Realization of efficient channel estimation using
  programmable metasurface,'' in \emph{2021 IEEE USNC-URSI Radio Sci. Meeting},
  Dec. 2021, pp. 66--67.

\bibitem{mandelli2023}
S.~Mandelli, M.~Henninger, and J.~Du, ``Sampling and reconstructing angular
  domains with uniform arrays,'' \emph{IEEE Trans. Wireless Commun.}, vol.~22,
  no.~6, pp. 3628--3642, Jun. 2023.

\bibitem{zhang2019}
J.~A. {Zhang}, X.~{Huang}, Y.~J. {Guo}, J.~{Yuan}, and R.~W. {Heath},
  ``Multibeam for joint communication and radar sensing using steerable analog
  antenna arrays,'' \emph{IEEE Trans. Veh. Technol.}, vol.~68, no.~1, pp.
  671--685, Jan. 2019.

\bibitem{liyanaarachchi2021_allocation}
S.~D. Liyanaarachchi, T.~Riihonen, C.~B. Barneto, and M.~Valkama, ``Optimized
  waveforms for {5G–6G} communication with sensing: Theory, simulations and
  experiments,'' \emph{IEEE Trans. Wireless Commun.}, vol.~20, no.~12, pp.
  8301--8315, Dec. 2021.

\bibitem{schmidl1997}
T.~M. {Schmidl} and D.~C. {Cox}, ``Robust frequency and timing synchronization
  for {OFDM},'' \emph{IEEE Trans. Commun.}, vol.~45, no.~12, pp. 1613--1621,
  Dec. 1997.

\bibitem{omri2019}
A.~Omri, M.~Shaqfeh, A.~Ali, and H.~Alnuweiri, ``Synchronization procedure in
  {5G} {NR} systems,'' \emph{IEEE Access}, vol.~7, pp. 41\,286--41\,295, Mar.
  2019.

\bibitem{bookSFO}
L.~Smaini, ``{RF} analog impairments description and modeling,'' in \emph{{RF}
  Analog Impairments Modeling for Communication Systems Simulation: Application
  to {OFDM}-based Transceivers}.\hskip 1em plus 0.5em minus 0.4em\relax
  Chichester, UK: Wiley, 2012, ch.~2, pp. 37--105.

\bibitem{chiueh2012}
T.-D. Chiueh, P.-Y. Tsai, and I.-W. Lai, ``Synchronization,'' in \emph{Baseband
  Receiver Design for Wireless MIMO-OFDM Communications, 2nd edition}.\hskip
  1em plus 0.5em minus 0.4em\relax Singapore: Wiley, 2012, ch.~6, pp. 127--165.

\bibitem{brunner2024}
D.~{Brunner et al.}, ``Bistatic {OFDM}-based {ISAC} with over-the-air
  synchronization: System concept and performance analysis,'' \emph{arXiv
  preprint arXiv:2405.04962 [eess.SP]}, May 2024.

\bibitem{hsieh1998}
M.-H. Hsieh and C.-H. Wei, ``Channel estimation for {OFDM} systems based on
  comb-type pilot arrangement in frequency selective fading channels,''
  \emph{IEEE Trans. Consum. Electron.}, vol.~44, no.~1, pp. 217--225, Feb.
  1998.

\bibitem{coleri2002}
S.~Coleri, M.~Ergen, A.~Puri, and A.~Bahai, ``Channel estimation techniques
  based on pilot arrangement in {OFDM} systems,'' \emph{IEEE Trans.
  Broadcast.}, vol.~48, no.~3, pp. 223--229, Sept. 2002.

\bibitem{kanhere2021_multistatic}
O.~Kanhere, S.~Goyal, M.~Beluri, and T.~S. Rappaport, ``Target localization
  using bistatic and multistatic radar with {5G} {NR} waveform,'' in \emph{2021
  IEEE 93rd VVeh. Technol. Conf.}, Apr. 2021, pp. 1--7.

\bibitem{dwivedi2021}
S.~{Dwivedi et al.}, ``Positioning in {5G} networks,'' \emph{IEEE Commun.
  Mag.}, vol.~59, no.~11, pp. 38--44, Nov. 2021.

\bibitem{wei2023}
Z.~{Wei et al.}, ``{5G} {PRS}-based sensing: A sensing reference signal
  approach for joint sensing and communication system,'' \emph{IEEE Trans. Veh.
  Technol.}, vol.~72, no.~3, pp. 3250--3263, Mar. 2023.

\bibitem{braun2011}
M.~Braun, C.~Sturm, and F.~K. Jondral, ``On the single-target accuracy of
  {OFDM} radar algorithms,'' in \emph{2011 IEEE 22nd Int. Symp. Pers., Indoor
  Mobile Radio Commun.}, Sept. 2011, pp. 794--798.

\bibitem{hwang2009}
T.~{Hwang}, C.~{Yang}, G.~{Wu}, S.~{Li}, and G.~{Ye Li}, ``{OFDM} and its
  wireless applications: A survey,'' \emph{IEEE Trans. on Veh. Technol.},
  vol.~58, no.~4, pp. 1673--1694, Aug. 2009.

\bibitem{wang2023}
L.~Wang, Z.~Wei, L.~Su, Z.~Feng, H.~Wu, and D.~Xue, ``Coherent compensation
  based {ISAC} signal processing for long-range sensing: (invited paper),'' in
  \emph{2023 21st Int. Symp. Model. Omptim. Mobile, Ad Hoc, Wireless Netw.},
  Aug. 2023, pp. 689--695.

\bibitem{erup1993}
L.~Erup, F.~Gardner, and R.~Harris, ``Interpolation in digital modems — part
  {II}: Implementation and performance,'' \emph{IEEE Trans. Commun.}, vol.~41,
  no.~6, pp. 998--1008, Jun. 1993.

\bibitem{bhutani2019}
A.~{Bhutani}, S.~{Marahrens}, M.~{Gehringer}, B.~{G\"ottel}, M.~{Pauli}, and
  T.~{Zwick}, ``The role of millimeter-waves in the distance measurement
  accuracy of an {FMCW} radar sensor,'' \emph{Sensors}, vol.~19, no.~18, pp.
  1--16, Sept. 2019.

\bibitem{li2021}
Y.~{Li et al.}, ``Beamsteering for {5G} mobile communication using programmable
  metasurface,'' \emph{IEEE Wireless Commun. Lett.}, vol.~10, no.~7, pp.
  1542--1546, Jul. 2021.

\bibitem{li2024}
------, ``User detection in {RIS}-based {mmWave} {JCAS}: Concept and
  demonstration,'' \emph{IEEE Trans. Wireless Commun. (Early Access)}, pp.
  1--16, Feb. 2024.

\bibitem{miao2024}
S.~{Miao et al.}, ``Trends in channel coding for {6G},'' \emph{Proc. IEEE
  (Early Access)}, pp. 1--23, 2024.

\bibitem{ETSI302307}
{ETSI}, ``{Digital Video Broadcasting (DVB); Second generation framing
  structure, channel coding and modulation systems for Broadcasting,
  Interactive Services, News Gathering and other broadband satellite
  applications; Part 1: DVB-S2},'' European Telecommunications Standards
  Institute (ETSI), TS 302 307-1 V1.4.1, Jul. 2014.

\bibitem{filomeno2022}
M.~{de Lima Filomeno et al.}, ``Joint channel estimation and {Schmidl} \& {Cox}
  synchronization for {OCDM}-based systems,'' \emph{IEEE Commun. Lett.},
  vol.~26, no.~8, pp. 1878--1882, Aug. 2022.

\bibitem{giroto2023_EuMW_PMCW}
L.~{Giroto de Oliveira et al.}, ``Enabling joint radar-communication operation
  in shift register-based {PMCW} radars,'' in \emph{2023 20th Eur. Radar
  Conf.}, Sept. 2023, pp. 85--88.

\end{thebibliography}

\end{document}